\documentclass{article}
\usepackage[preprint]{log_2023}
\usepackage{multirow}
\usepackage{xspace}
\usepackage{soul}
\usepackage{pgfplots}
\usepackage{booktabs}   
\usepackage{tcolorbox}
\usepackage{pifont}% http://ctan.org/pkg/pifont
\usepackage{makecell}
\usepackage{adjustbox}
\usepackage{multirow}            % tabular cells spanning multiple rows
\usepackage{amsfonts}            % blackboard math symbols
\usepackage{graphicx}            % figures
\usepackage{duckuments} 
\usepackage[ruled,vlined]{algorithm2e}

\pgfplotsset{compat=1.18}

\definecolor{gray}{rgb}{0.83, 0.83, 0.83}

\newcommand{\tikzcircle}[2][red,fill=red]{\tikz[baseline=-0.5ex]\draw[#1,radius=#2] (0,0) circle ;}

\usepackage{geometry}
\usepackage{array}
\usepackage{xcolor}

\usepackage{color, colortbl}
	
\definecolor{Gray}{gray}{0.9}

\usepackage[numbers,compress,sort]{natbib}

\title{A Topology-aware Analysis of Graph Collaborative Filtering}

\author{%
Daniele Malitesta\\
Politecnico di Bari\\
\email{daniele.malitesta@poliba.it}\And
Claudio Pomo\\
Politecnico di Bari\\
\email{claudio.pomo@poliba.it}\And
Vito W. Anelli\\
Politecnico di Bari\\
\email{vitowalter.anelli@poliba.it}\And
Alberto C. M. Mancino\\
Politecnico di Bari\\
\email{alberto.mancino@poliba.it}\And
Eugenio {Di Sciascio}\\
Politecnico di Bari\\
\email{eugenio.disciascio@poliba.it}\And
Tommaso {Di Noia}\\
Politecnico di Bari\\
\email{tommaso.dinoia@poliba.it}
}

\begin{document}

\maketitle

\begin{abstract}
The successful integration of graph neural networks into recommender systems (RSs) has led to a novel paradigm in collaborative filtering (CF), graph collaborative filtering (graph CF). By representing user-item data as an undirected, bipartite graph, graph CF utilizes short- and long-range connections to extract collaborative signals that yield more accurate user preferences than traditional CF methods. Although the recent literature highlights the efficacy of various algorithmic strategies in graph CF, the impact of datasets and their topological features on recommendation performance is yet to be studied. To fill this gap, we propose a topology-aware analysis of graph CF. In this study, we (i) take some widely-adopted recommendation datasets and use them to generate a large set of synthetic sub-datasets through two state-of-the-art graph sampling methods, (ii) measure eleven of their classical and topological characteristics, and (iii) estimate the accuracy calculated on the generated sub-datasets considering four popular and recent graph-based RSs (i.e., LightGCN, DGCF, UltraGCN, and SVD-GCN). Finally, the investigation presents an explanatory framework that reveals the linear relationships between characteristics and accuracy measures. The results, statistically validated under different graph sampling settings, confirm the existence of solid dependencies between topological characteristics and accuracy in the graph-based recommendation, offering a new perspective on graph CF.
\end{abstract}

\section{Introduction and Motivation}

\noindent \textbf{Context.} Collaborative filtering~\cite{DBLP:journals/fthci/EkstrandRK11} (CF) is among the prime strategy patterns in personalized recommendation. Recommender systems (RSs) following the CF rationale assume that users having enjoyed similar items in the past could be captivated to interact with similar items also in the near future. Such a paradigm usually maps users and items to embeddings in a latent space and optimizes an objective function to derive hidden patterns from the historical preference data~\cite{DBLP:journals/computer/KorenBV09, DBLP:conf/uai/RendleFGS09, DBLP:conf/www/HeLZNHC17}.

Recently, models built upon graph neural networks~\cite{DBLP:series/synthesis/2020Hamilton} (GNNs), such as graph convolutional networks~\cite{DBLP:conf/iclr/KipfW17} (GCNs) have provided a novel outlook on the traditional CF approaches~\cite{DBLP:journals/csur/WuSZXC23}. By viewing the user-item preference data as a bipartite and undirected graph, GCN-based techniques have shown the potential to mine near- and long-distance preferences of users towards items. Differently from the classical CF models~\cite{DBLP:conf/uai/RendleFGS09, DBLP:conf/www/HeLZNHC17}, which utilize the information from user-item interactions only in the optimization of their objective functions, graph convolution is designed to encapsulate user-item multi-hop relationships right from the representation of their embeddings. Such learned profiles are suitably exploited to produce more accurate recommendations~\cite{DBLP:conf/cikm/MaoZXLWH21, DBLP:conf/sigir/ZhuDSMLCXZ22}. 

\noindent \textbf{Related work.} The vanilla GCN layer works by performing the message-passing schema, which iteratively refines users' and items' node representations with the aggregation of their multi-hop neighbor nodes. After early attempts~\cite{DBLP:journals/corr/BergKW17}, more sophisticated solutions proposed to exploit the inter-dependencies among nodes and neighbors~\cite{DBLP:conf/sigir/Wang0WFC19}, lighten the procedure for the message-passing~\cite{DBLP:conf/sigir/0001DWLZ020}, and learn multiple nodes' views~\cite{DBLP:conf/sigir/WuWF0CLX21, DBLP:conf/sigir/YuY00CN22, DBLP:conf/sigir/0001OM22, DBLP:conf/iclr/Cai0XR23} through self-supervised~\cite{DBLP:conf/cikm/HuangXW0Y22} and contrastive~\cite{DBLP:conf/nips/KhoslaTWSTIMLK20} learning. Some recent trends suggest simplifying the usual formulation for the message-passing~\cite{DBLP:conf/cikm/MaoZXLWH21, DBLP:conf/cikm/PengSM22}, translating the graph-based recommendation task to other spaces~\cite{DBLP:conf/cikm/ShenWZSZLL21, DBLP:conf/www/SunCZPV21, DBLP:conf/sigir/ZhangL0WSLSZDZ22, DBLP:conf/icml/YuQ20}, and exploiting hyper-graphs to capture more complex user-item dependencies~\cite{DBLP:conf/sigir/XiaHXZYH22, DBLP:conf/cikm/WeiLBL22}. To filter out the noisy neighbor contributions and disclose (or explain~\cite{DBLP:conf/icml/Ma0KW019}) hidden preference patterns within the graph, a complementary research path was proposed to learn importance weights through attention mechanisms as in the graph attention network~\cite{DBLP:conf/iclr/VelickovicCCRLB18} (GAT). While some models are trained to recognize meaningful interactions on the user-item level~\cite{DBLP:conf/kdd/Wang00LC19, DBLP:journals/ipm/TaoWWHHC20, DBLP:conf/sigir/TianXLYZ22}, other ones disentangle such relations on a finer-grained scale~\cite{DBLP:conf/sigir/WangJZ0XC20, DBLP:conf/sigir/ZhangL0WSLSZDZ22}.

\noindent \textbf{Motivations.} From an \textit{algorithmic} perspective, the technical contribution of graph-based RSs has been theoretically~\cite{DBLP:conf/cikm/ShenWZSZLL21} and empirically~\cite{DBLP:conf/wsdm/WangZS22} investigated to justify their high-quality recommendations. On the one hand, established approaches such as LightGCN~\cite{DBLP:conf/sigir/0001DWLZ020} and DGCF~\cite{DBLP:conf/sigir/WangJZ0XC20} re-adapt the GCN layer to suit the collaborative filtering schema. Specifically, the former suggests that feature transformations and non-linearities should be removed since they could negatively impact the recommendation performance; the latter recognizes the importance of updating the user-item graph structure according to the learned users' intents towards items. On the other hand, recent graph-based RSs including UltraGCN~\cite{DBLP:conf/cikm/MaoZXLWH21} and SVD-GCN~\cite{DBLP:conf/cikm/PengSM22} acknowledge that existing graph-based models may still be affected by the over-smoothing phenomenon and scalability issues. 
To overcome such problems, they propose to go beyond the traditional concept of message aggregation at multiple layers by adopting ad-hoc mathematical proxies of the message-passing. UltraGCN approximates infinite propagation layers, while SVD-GCN explores the analogies between graph convolution and SVD. Additionally, both approaches learn from user-user and item-item relationships.

From another perspective, the \textit{machine learning} literature acknowledges that graph \textit{topology} plays a crucial role in GNNs. The authors in \cite{DBLP:conf/www/WeiZH22} recognize how topology may increase the model's capacity by distinguishing between two approaches that influence the (topological) shape of GNNs. The first approach involves stacking aggregation operations, while the second one utilizes multiple aggregations of operations to prevent over-smoothing and enhance model capacity. Indeed, the learning ability of GNNs is a delicate balance between topology and node attributes. While the graph convolutional layer is gaining momentum due to Laplacian smoothing or low-pass filtering~\cite{DBLP:conf/icml/WuSZFYW19}, stacking too many such layers can lead to a loss of expressive power, causing node representations to become dependent solely on node degree and connectivity. Two strategies are under debate to alleviate this issue~\cite{DBLP:conf/www/0002ZPNG0CH22}: modifying the learned representations~\cite{DBLP:conf/iclr/KlicperaBG19,DBLP:conf/iclr/ZhaoA20} and modifying network topology~\cite{DBLP:conf/icml/Abu-El-HaijaPKA19,DBLP:conf/iclr/KlicperaBG19,DBLP:conf/nips/LuanZCP19,DBLP:conf/icml/XuLTSKJ18,DBLP:conf/iclr/RongHXH20}. In summary, the inconsistency between network topology and node content exists at the individual node and network levels. A recent study~\cite{DBLP:journals/tnn/ShiTZ22} suggests revisiting network topology to enhance GCN learning, but this involves revisiting the network topology thus modifying the underlying method.

In recommendation, some works have demonstrated how \textit{classical} dataset statistics may impact the performance of recommendation models~\cite{DBLP:journals/tmis/AdomaviciusZ12,DBLP:conf/sigir/DeldjooNSM20,DBLP:conf/wsdm/SachdevaWM22}, for example, data sparsity within the user-item interaction matrix, the concentration of interactions from both user and item viewpoints or the ratio of users to items comprising the catalog. However, we believe that the topological nature of the user-item data (which allows us to interpret users/items and their interactions as a bipartite and undirected graph) along with the novel graph CF wave, could require a more careful analysis of additional (and less shallow) \textit{topological} measures. By describing the topology of the user-item graph under the lens of its (multi)-hop connections, we regard it as imperative to unravel the interdependencies between topology-aware attributes and recommendation performance to better understand graph CF. 

\noindent \textbf{Our contributions.} Motivated by the reasons above, we propose a topology-aware analysis of graph collaborative filtering to find the (possible) dependencies among (topological) data characteristics and recommendation performance of graph-based recommender systems. To this end, we first consider two popular datasets (i.e., Yelp2018 and Gowalla) and generate 1,200 synthetic sub-datasets with two common graph sampling strategies (i.e., node- and edge-dropout). Second, we select eleven \textit{classical} and \textit{topological} characteristics which are loosely correlated with one another and measure them for each of the sampled sub-datasets\footnote{We indicate with \textit{classical} those recommendation-based measures exploited in other similar works~\cite{DBLP:journals/tmis/AdomaviciusZ12, DBLP:conf/sigir/DeldjooNSM20}, and with \textit{topological} those properties of the user-item graph conceptually related to node degree~\cite{DBLP:journals/socnet/LatapyMV08, PhysRevE.67.026126}.}. Then, we choose four graph-based RSs, namely, LightGCN~\cite{DBLP:conf/sigir/0001DWLZ020}, DGCF~\cite{DBLP:conf/sigir/WangJZ0XC20}, UltraGCN~\cite{DBLP:conf/cikm/MaoZXLWH21}, and SVD-GCN~\cite{DBLP:conf/cikm/PengSM22}, since they are state-of-the-art approaches in the related literature. Our contributions may be summarized as follows:
%: (i) they are recent approaches, (ii) they are used as baselines in several works from the last years, (iii) they adopt different strategies that well depict most of the models from the literature. Last, an explanatory framework is trained to identify linear relations among characteristics and accuracy metrics. Our contributions are:
\begin{enumerate}
    \item To the best of our knowledge, this is the first work that analyzes the influence of \textit{classical} and \textit{topological} dataset characteristics on the performance of state-of-the-art graph-based RSs, with a re-interpretation of the topology-aware characteristics under the lens of recommender systems.
    \item We carefully choose four graph-based RSs that are across-the-board and recently proposed in the literature and also span an extensive selection of graph strategies. In an attempt to make the information conveyed by node degree explicitly emerge from their formulations, we aim to understand how each of them addresses such a topological aspect in all its facets.
    \item We build an explanatory framework to calculate the linear dependencies among characteristics (the independent variables) and recommendation metrics (the dependent variables).
    \item We validate the proposed framework on its statistical significance and uncover insights on graph CF under the novel perspective of graph topology. We further test the efficacy of the approach with varying settings of graph samplings, shedding light on the influence of node- and edge-dropout for our explanatory model.
\end{enumerate}

Code and datasets to reproduce all results are available at:~\url{https://github.com/sisinflab/Graph-Characteristics}.
\section{Topological Characteristics in Recommendation Data}
\label{sec:topological-characteristics}

By viewing the recommendation data as a bipartite and undirected user-item graph, we describe its \textit{topological} characteristics~\cite{DBLP:journals/socnet/LatapyMV08, PhysRevE.67.026126}, which we re-interpret from the viewpoint of RSs.

\subsection{Preliminaries}

In a recommendation system, we denote with $\mathcal{U}$ and $\mathcal{I}$ the sets of users and items, respectively, where $|\mathcal{U}| = U$ and $|\mathcal{I}| = I$. Then, we indicate with $\mathbf{R} \in \mathbb{R}^{U \times I}$ the interaction matrix collecting user-item interactions in the form of implicit feedback (i.e., $\mathbf{R}_{u, i} = 1$ if user $u \in \mathcal{U}$ interacted with item $i \in \mathcal{I}$, 0 otherwise). Moreover, let $\mathcal{N}_u = \{i\;|\; \mathbf{R}_{u, i} = 1\}$ and $\mathcal{N}_i = \{u\;|\;\mathbf{R}_{u, i} = 1\}$ be the sets of items and users having an interaction with $u$ and $i$, respectively. We use $\mathbf{R}$ to define the adjacency matrix $\mathbf{A} \in \mathbb{R}^{(U + I) \times {(U + I)}}$ representing the bidirectional interactions between users and items:
\begin{equation}
    \mathbf{A} = \begin{bmatrix}
    0 & \mathbf{R} \\
    \mathbf{R}^\top & 0 
    \end{bmatrix}.
\end{equation}
On such basis, let $\mathcal{G} = \{\mathcal{U} \cup \mathcal{I}, \mathbf{A}\}$ be the user-item bipartite and undirected graph. Moreover, we connote the user- and item-\textit{projected} graphs as $\mathcal{G}_\mathcal{U} = \{\mathcal{U}, \mathbf{A}^{\mathcal{U}}\}$ and $\mathcal{G}_\mathcal{I} = \{\mathcal{I}, \mathbf{A}^{\mathcal{I}}\}$. Thus, let $\mathbf{R}^{\mathcal{U}}$ and $\mathbf{R}^{\mathcal{I}}$ be the user-user and item-item interaction matrices:
\begin{equation}
    \mathbf{R}^{\mathcal{U}} = \mathbf{R} \cdot \mathbf{R}^\top, \qquad \mathbf{R}^{\mathcal{I}} = \mathbf{R}^\top \cdot \mathbf{R},
\end{equation}
which indicate the co-occurrences among users and items, respectively. Trivially, the corresponding adjacency matrices $\mathbf{A}^{\mathcal{U}}$ and $\mathbf{A}^{\mathcal{I}}$ are:
\begin{equation}
    \mathbf{A}^{\mathcal{U}} = \mathbf{R}^{\mathcal{U}}, \qquad \mathbf{A}^{\mathcal{I}} = \mathbf{R}^{\mathcal{I}}.
\end{equation}
We use the introduced concepts and notations to describe three topological aspects of the user-item graph and re-interpret them under the lens of recommender systems.

\subsection{Node degree} By generalizing the definitions of $\mathcal{N}_u$ and $\mathcal{N}_i$, let $\mathcal{N}^{(l)}_u$ and $\mathcal{N}^{(l)}_i$ be the sets of neighborhood nodes for user $u$ and item $i$ at $l$ distance hops. Thus, the node degrees for $u$ and $i$ (i.e., $\sigma_u = |\mathcal{N}^{(1)}_u|$ and $\sigma_i = |\mathcal{N}^{(1)}_i|$) represent the number of item and user nodes directly connected with $u$ and $i$, respectively. The average user and item node degrees are:
\begin{equation}
    \sigma_\mathcal{U} = \frac{1}{U} \sum_{u \in \mathcal{U}} |\mathcal{N}^{(1)}_u|, \qquad \sigma_\mathcal{I} = \frac{1}{I} \sum_{i \in \mathcal{I}} |\mathcal{N}^{(1)}_i|.
\end{equation}

\noindent\textsc{\bfseries RecSys re-interpretation.} \textit{The node degree in the user-item graph stands for the number of items (users) interacted by a user (item). This is related to the cold-start issue in recommendation, where cold users denote low activity on the platform, while cold items are niche products.}

Node degree alone still fails to provide a deeper outlook on the user-item graph. The following topology-aware characteristics, derived from node degree, expand its formulation to other viewpoints.

\subsection{Clustering coefficient} For each partition in a bipartite graph, it is interesting to recognize clusters of nodes in terms of how their neighborhoods overlap, independently of the respective sizes. Let $v$ and $w$ be two nodes from the same partition (e.g., user nodes). Their similarity is the intersection over union of their neighborhoods~\cite{DBLP:journals/socnet/LatapyMV08}. By evaluating the metric node-wise, we obtain:
\begin{equation}
    \gamma_{v} = \frac{\sum_{w \in \mathcal{N}^{(2)}_v}\gamma_{v, w}}{|\mathcal{N}_v^{(2)}|}, \qquad \text{with} \quad  \gamma_{v, w} = \frac{|\mathcal{N}^{(1)}_{v} \cap \mathcal{N}^{(1)}_{w}|}{|\mathcal{N}^{(1)}_{v} \cup \mathcal{N}^{(1)}_{w}|},
\end{equation}
where $\mathcal{N}_v^{(2)}$ is the second-order neighborhood set of $v$. In this case, we leverage the second-order neighborhood because, in a bipartite graph, nodes from the same partition are connected at (multiple of) 2 hops. The average clustering coefficient on $\mathcal{U}$ and $\mathcal{I}$ is:
\begin{equation}
    \gamma_{\mathcal{U}} = \frac{1}{U} \sum_{u \in \mathcal{U}} \gamma_u, \qquad \gamma_{\mathcal{I}} = \frac{1}{I} \sum_{i \in \mathcal{I}} \gamma_i.
\end{equation}

\noindent\textsc{\bfseries RecSys re-interpretation.} \textit{High values of the clustering coefficient indicate that there exists a substantial number of co-occurrences among nodes from the same partition. For instance, when considering the user-side formula, the average clustering coefficient increases if several users share most of their interacted items. The intuition aligns with the rationale behind collaborative filtering: two users are likely to show similar preferences when they interact with the same items.}

The clustering coefficient allows the description of broader portions of the user-item graph compared to the semantics conveyed by node degree. Indeed, the measure takes nodes at 2 hops (i.e., user-item-user and item-user-item connections). Nevertheless, we may want to capture properties for even more extended regions of the graph. For this reason, we introduce one last topology-aware characteristic that goes beyond the 2-hop distance among nodes.   

\subsection{Degree assortativity}
\label{sec:assortativity} In real-world graphs, nodes tend to gather when they share similar characteristics. Such a tendency is measured through the assortativity coefficient. Depending on the semantics of ``node similarity'', there exist different formulations for assortativity~\cite{PhysRevE.67.026126}. For the sake of this work, we consider the assortativity coefficient based on the scalar properties of graph nodes, for instance, their degree. Let $\mathcal{D} = \{d_1, d_2, \dots\}$ be the set of unique node degrees in the graph, and let $e_{d_h, d_k}$ be the fraction of edges connecting nodes with degrees $d_h$ and $d_k$. Then, let $q_{d_h}$ be the probability distribution to choose a node with degree $d_h$ after having selected a node with the same degree (i.e., the \textit{excess} degree distribution). The degree assortativity coefficient is calculated as:
\begin{equation}
    \rho = \frac{\sum\limits_{d_h, d_k}d_h d_k(e_{d_h, d_k} - q_{d_h} q_{d_k})}{std^2_q},
\end{equation}
where $std_q$ is the standard deviation of the distribution $q$. Note that, for its formulation, the degree assortativity is similar to a correlation measure (e.g., Pearson correlation). Following the same rationale of the clustering coefficient, we are interested in finding similarity patterns among nodes from the same partition. For this reason, we first apply the projection of the user-item bipartite graph for both users and items to obtain the user- (i.e., $\mathcal{G}_{\mathcal{U}}$) and item- (i.e., $\mathcal{G}_{\mathcal{I}}$) projected graphs. Then, we calculate the degree assortativity coefficients for $\mathcal{G}_{\mathcal{U}}$ and $\mathcal{G}_{\mathcal{I}}$, namely, $\rho_{\mathcal{U}}$ and $\rho_{\mathcal{I}}$.

\noindent\textsc{\bfseries RecSys re-interpretation.} \textit{In the recommendation scenario, the degree assortativity calculated user- and item-wise is a proxy to represent the tendency of users with the same activity level on the platform and items with the same popularity to gather, respectively. Since we calculate the degree assortativity on the complete user-user and item-item co-occurrence graphs, we deem this characteristic to provide a broader view of the dataset than the clustering coefficient. For this reason, to give an intuition of degree assortativity, we borrow the concept of search space traversal depth in search algorithms theory. That is, we re-interpret degree assortativity in recommendation as a topology-aware characteristic showing a strong look-ahead nature.}

To conclude this section, we suggest the reader refer to Appendix \ref{app:class_charact} where we also recall the \textit{classical} characteristics underlying the user-item data as presented in~\cite{DBLP:journals/tmis/AdomaviciusZ12, DBLP:conf/sigir/DeldjooNSM20}.
\section{Topological Characteristics in Graph Collaborative Filtering}
\label{sec:top-char-graph-collab}

Since graph-based recommender systems are specifically designed to view the user-item interaction data as a bipartite and undirected graph, in this section, we seek to understand how and to what extent such models (explicitly) integrate \textit{topological} data characteristics into their formulations. To this aim, we select four popular and recent approaches in graph collaborative filtering, namely: LightGCN~\cite{DBLP:conf/sigir/0001DWLZ020}, DGCF~\cite{DBLP:conf/sigir/WangJZ0XC20}, UltraGCN~\cite{DBLP:conf/cikm/MaoZXLWH21}, and SVD-GCN~\cite{DBLP:conf/cikm/PengSM22}.

Table \ref{tab:models} shows how such techniques are widely used as baselines in the recent literature and indicates which \textit{topological} characteristics are explicitly involved in their formulations. While the interested reader may refer to Appendix \ref{app:graph-recsys} for a more extended theoretical study, we observe that:

\noindent\textsc{\bfseries Observation.} \textit{The analyzed graph-based recommender systems explicitly utilize the node degree information during the representation learning phase, each of them in a different way. However, clustering coefficient and degree assortativity, which share similarities with node degree's semantics, do not seem to have an evident representation within the models' formulations. Under this perspective, our study will also serve to test what topological aspects graph-based RSs can (un)intentionally capture during their training.}

Indeed, these observations pave the way to a further question: \textit{\textbf{are (topological) dataset characteristics influencing the recommendation performance of graph-based recommender systems?}}
\section{Proposed Analysis}
To answer such a question, in the following, we present our proposed analysis to evaluate the impact of \textit{classical} and \textit{topological} characteristics on the performance of graph-based recommender systems. As already done in similar works~\cite{DBLP:journals/tmis/AdomaviciusZ12, DBLP:conf/sigir/DeldjooNSM20}, we decide to design an explanatory statistical model which finds dependencies between dataset characteristics and recommendation performance. In this respect, it becomes imperative to collect a set of samples (with dataset characteristics and the corresponding models' recommendation performance) that is large enough to ensure the statistical significance of the conducted analysis under a certain confidence threshold. Thus, in the following, we first present the approach to generate an extensive set of sub-datasets from two popular recommendation datasets, which represent our set of samples. Then, we specify and justify the design choices for the explanatory model which we adopt in our analysis to fit the dataset characteristics of the sub-datasets to the recommendation performance of graph-based recommender systems measured on the same sub-datasets.

\subsection{Dataset generation}~\label{dataset-generation}

The literature has recently demonstrated that dataset sampling in collaborative filtering can robustify the training of recommendation models~\cite{DBLP:conf/sigir/WuWF0CLX21}, sometimes with ad-hoc solutions performed end-to-end in the downstream task~\cite{DBLP:conf/kdd/ChenSSH17, DBLP:conf/wsdm/SachdevaWM22}. In this work, we propose to adopt such strategies \textit{\textbf{for another purpose}}, namely, the random generation of synthetic (but meaningful) data to conduct our study. We adopt a similar approach to other studies that examine how \textit{classical} characteristics affect the performance of recommendation models~\cite{DBLP:journals/tmis/AdomaviciusZ12, DBLP:conf/sigir/DeldjooNSM20}. However, we take it a step further by analyzing these aspects from a \textit{topological} perspective. For this reason, we use node- and edge-dropout strategies to generate the sub-datasets, which have gained recent attention in graph learning literature~\cite{DBLP:conf/sigir/WuWF0CLX21, DBLP:conf/www/ShuXLWKM22}. Since the goal is to generate sub-datasets exhibiting maximum \textit{topological} diversity, we depart from the conventional method of utilizing a Bernoulli distribution to randomly perturb the adjacency matrix. Instead, we vary the level of aggressiveness (i.e., dropout rate) that involves sampling with node- or edge-dropout. Further details about the two distinct \textit{graph sampling strategies} and \textit{sub-dataset generation} are presented in the following paragraphs.

\begin{table}[!t]
\centering
\caption{Selected models for our study. For each of them, we report year, works using them as baselines, and which \textit{topological} characteristics are integrated in the models' formulations.}
\footnotesize
\setlength{\tabcolsep}{10pt} % Default value: 6pt
\renewcommand{\arraystretch}{1.5} % Default value: 1
\begin{tabular}{lccl}
\toprule
\multicolumn{1}{c}{\textbf{Model}} & \multicolumn{1}{c}{\textbf{Year}} & \multicolumn{1}{c}{\textbf{Baseline in (2021-)}} & \multicolumn{1}{c}{\textbf{Topological characteristics}} \\ \hline
LightGCN~\cite{DBLP:conf/sigir/0001DWLZ020} & 2020 & e.g.,~\cite{DBLP:conf/sigir/WuWF0CLX21, DBLP:conf/www/LiuCZGN21, DBLP:conf/sigir/YuY00CN22, DBLP:conf/sigir/XiaHXZYH22, DBLP:conf/kdd/RaoCLSYH22} & 
\parbox[t]{5cm}{\textbf{Node degree} used to normalize the adjacency matrix in the message passing.} \\ \hline
DGCF~\cite{DBLP:conf/sigir/WangJZ0XC20} & 2020 & e.g.,~\cite{ DBLP:conf/sigir/ZhangL0WSLSZDZ22, DBLP:conf/sigir/FanL0ZT022, 9736612, DBLP:conf/wsdm/WangZS22, DBLP:conf/www/LinTHZ22} & \parbox[t]{5cm}{\textbf{Node degree} used to normalize the adjacency matrix in the message passing.} \\ \hline
UltraGCN~\cite{DBLP:conf/cikm/MaoZXLWH21} & 2021 & e.g.,~\cite{DBLP:conf/mm/DuWF0022, DBLP:conf/cikm/GongSWLL22, DBLP:conf/sigir/ZhuDSMLCXZ22,DBLP:conf/sigir/0001OM22, 10.1145/3570501} & \parbox[t]{5cm}{\textbf{Node degree} used for normalization in the infinite layer message passing. The model also learns from the \textbf{\textit{item}-projected graph}.} \\ \hline
SVD-GCN~\cite{DBLP:conf/cikm/PengSM22} & 2022 & e.g.,~\cite{DBLP:journals/corr/abs-2306-02259, DBLP:journals/corr/abs-2306-04370} & \parbox[t]{5cm}{The formulation for the node embeddings involves the largest singular values of the normalized user-item interaction matrix, whose maximum value is related to the maximum \textbf{node degree} of the user-item graph. The model also learns from the \textbf{\textit{user}-} and \textbf{\textit{item}-projected graphs}.} \\
 \bottomrule
 \end{tabular}
\label{tab:models}
\end{table}

\noindent\textbf{Graph sampling.} Given the bipartite user-item graph $\mathcal{G}$, a dropout rate $\mu$, and a \texttt{\textit{sampling}} strategy, the algorithm returns the sampled graph $\mathcal{G}_m$. When the \texttt{\textit{sampling}} strategy is \texttt{nodeDropout}, the procedure initially calculates the number of nodes to sample from the graph and uniformly extracts the new set of nodes (i.e., $\mathcal{V}_m$). Then, the adjacency matrix is masked to obtain a new one with the retained nodes only (i.e., $\mathbf{A}_m$). Conversely, when the \texttt{\textit{sampling}} strategy is \texttt{edgeDropout}, the procedure initially calculates the number of edges to sample from the graph and uniformly extracts the new set of edges (i.e., $\mathcal{E}_m$). Then, the adjacency matrix is masked to obtain a new one with the retained edges only (i.e., $\mathbf{A}_m$) and the corresponding set of nodes (i.e., $\mathcal{V}_m$). Finally, in both cases, the graph $\mathcal{G}_m$ is induced through $\mathcal{V}_m$ and $\mathbf{A}_m$.

\noindent\textbf{Sub-dataset generation.} The procedure takes the bipartite user-item graph $\mathcal{G}$ and the number of samples $M$ as inputs. Iterating for $M$ times, each sampled graph $\mathcal{G}_m$ is generated (through the previous algorithm) and added to the set of sub-datasets (i.e., $\mathcal{M}$). Specifically, during each iteration, a dropout rate $\mu$ is uniformly sampled from the range $[0.7, 0.9]$ to ensure the generation of small sub-samples of the original dataset. Then, either \texttt{nodeDropout} or \texttt{edgeDropout} is chosen at random, thus the overall procedure is not biased towards one of the \texttt{\textit{sampling}} strategies (see also Section \ref{sec:rq2}). Finally, the graph $\mathcal{G}_m$ is obtained by performing the selected \texttt{\textit{sampling}} strategy. 

The reader may refer to Appendix \ref{app:alg} for the pseudocode of both algorithms.

\subsection{Explanatory model}\label{sec:explanatory_model}
%Besides prediction, statistical models are useful tools to provide explanations. Such models unveil how certain factors (independent variables) influence a measure (dependent variable). In other words, 
% Statistical models may be used to supply an interpretation regarding the relationship between a hypothetical cause of a phenomenon (the independent variables) measured through the dependent variables. The current study employs statistical models to identify the dependencies between the characteristics of a dataset to the recommendation performance. For this purpose, the literature demonstrates that several potential functions could be used to fit the independent variables to the dependent ones. However, we decide to exploit a linear regression model to: (i) follow the same methodology from recent studies~\cite{DBLP:journals/tmis/AdomaviciusZ12,DBLP:conf/sigir/DeldjooNSM20}, and (ii) derive explanations on the performance impact of data characteristics through linear dependencies (i.e., the most straightforward and intuitive strategy).
Statistical models can be utilized to elucidate the relationship between a hypothesized cause of a phenomenon (i.e., independent variables) and its effect (measured through dependent variables). While various potential functions can be used to fit the independent variables to the dependent ones, we opt to utilize a linear regression model for two reasons: (i) to adhere to the same methodology employed in recent studies such as~\cite{DBLP:journals/tmis/AdomaviciusZ12,DBLP:conf/sigir/DeldjooNSM20}, and (ii) to derive explanations on the performance impact of data characteristics through linear dependencies, which represents the most straightforward and intuitive strategy. Following this intuition, we formalize a regression model:
\begin{equation}
\mathbf{y}=\boldsymbol{\epsilon}+\theta_0+\boldsymbol{\theta}_c \mathbf{X}_c.
\label{base_formula}
\end{equation}
We recall that our goal is to test if the factors related to the data characteristics (i.e., $\mathbf{X}_{c}$) can explain the effect on the recommendation system's performance (i.e., $\mathbf{y}$). Therefore, in Equation \ref{base_formula}, we denote by $\boldsymbol{\theta}_c=\left[\theta_1, \ldots, \theta_C\right]$ the vector of regression coefficients each of whom is associated with the $c$-th feature (data characteristic considered here), $\mathbf{X}_c \in \mathbb{R}^{M \times C}$ the matrix containing the data characteristic values for each sample in the training set, and $\mathbf{y}$ the vector containing the values of the performance measure associated with all samples in the training set. Moreover, under the assumption of mean-centered data, $\theta_0$ expresses the expected value of $\mathbf{y}$ (i.e., in this case, the expected recommendation performance). The regression model is trained through Ordinary Least Squares (OLS):
\begin{equation}
\label{eq:OLS}
\left(\theta_0^*, \boldsymbol{\theta}_c^*\right)=\min _{\theta_0, \boldsymbol{\theta}_c} \frac{1}{2}\left\|\mathbf{y}-\theta_0-\boldsymbol{\theta}_c    \mathbf{X}_c\right\|_2^2.
\end{equation}
% We employ the simple regression model in Equation \ref{eq:OLS} to maximize the $R^2$ value. This enables us to express the influence of the $\mathbf{X}_{c}$ coefficients on the recommendation system’s performance~\cite{gareth2013introduction}.
In order to show how the performance of RSs are related to dataset characteristics, we utilize the basic regression model presented in Equation~\ref{eq:OLS} with the aim of maximizing the $R^2$ coefficient. This approach allows us to effectively motivate the impact of the $\boldsymbol{\theta}_{c}$ coefficients on the recommendation system's effectiveness, as outlined in~\cite{gareth2013introduction} for any regression model.
\section{Results and Discussion}

This section aims to test the effectiveness of our proposed explanatory model. Specifically, we seek to answer the following two research questions: \textbf{RQ1)} What is the impact of classical and topological characteristics on the performance of graph-based recommender systems?; \textbf{RQ2)} Is the generation of sub-datasets through node- and edge-dropout differently influencing the explanations of our model? The reader may refer to Appendix \ref{app:exp_sett} for a detailed presentation of the experimental settings. 

\subsection{Impact of characteristics (RQ1)}
We assess the impact of \textit{classical} and \textit{topological} characteristics on the accuracy performance (i.e., \textit{Recall@20}) of graph-based RSs. Table \ref{tab:rq1} displays the results for our proposed explanatory model. In the first row (in light gray) we evaluate the goodness of the explanatory model through the $R^2$ and its adjusted version denoted as adj. $R^2$~\cite{DBLP:journals/tmis/AdomaviciusZ12,DBLP:conf/sigir/DeldjooNSM20}. Conversely, the remaining rows show the learned characteristics' coefficients (i.e., $[\theta_1, \dots, \theta_C]$ as described in Section \ref{sec:explanatory_model}, and renamed through a more human-readable convention as reported in Appendix \ref{app:char-calc}). Trivially, coefficients' signs and values indicate whether there exists a (strong) direct/inverse relation between the recommendation metric and the dataset characteristic. Finally, we assess the statistical significance of the results (the asterisks alongside each characteristic's coefficient, refer to the legend below the table). The reader may refer to Appendix \ref{app:add_exp_1} for additional experiments performed for RQ1.

Overall, the adj. $R^2$ is, for the vast majority of settings, above 95\%, proving the ability of the regression model to explain the accuracy recommendation performance through the measured characteristics. Hereinafter, we further decompose the regression results by categorizing the characteristics as \textit{classical} and \textit{topological}.

\noindent \textbf{Classical dataset characteristics.} 
Previous works~\cite{DBLP:journals/tmis/AdomaviciusZ12} have assessed the impact of \textit{classical} characteristics on neighbor- and factorization-based recommendation models.
However, a careful search of the relevant literature yields that no study has investigated whether such characteristics influence graph-based recommender systems likewise. 
From a theoretical standpoint, it could be noticed that neighbor- and factorization-based models have some similarities with graph-based ones, as the latter use a message-passing schema that aggregates information from the \textit{neighborhood} and also learn latent \textit{factor} representations of users and items.

Nevertheless, and interestingly, Table \ref{tab:rq1} suggests that the factorization component might be predominant within graph-based recommender systems. If we refer to the results from~\cite{DBLP:journals/tmis/AdomaviciusZ12}, we observe that factorization- and graph-based approaches are particularly aligned, considering: (i) the inverse correspondence between the accuracy performance metric and the $Shape_{log}$ in almost all settings, meaning that when the number of users is higher than the number of items in the system, accuracy performance may decrease; (ii) the direct correspondence between the accuracy performance metric and $Density_{log}$ and $Gini\text{-}I$. As for (ii), the density is historically known as one of the core problems in recommendation (i.e., data sparsity), so it becomes evident why also graph-based recommender systems' performance benefits from denser (i.e., less sparse) datasets. The Gini index measures the dissimilar distribution of items' interactions in the system and could be related to the tendency of RSs to promote popular items from the catalog. That is, when there exist items that have been experienced more frequently than others, both graph- and factorization-based RSs may be biased to popular items, and so their accuracy performance increases. Noteworthy, all such observations are supported by the statistical significance of the results.

\begin{table}[!t]
\caption{Results of the explanatory model with the \textit{Recall@20} as recommendation metric. Besides the row in light gray standing for the $R^2$, the other rows refer to the learned characteristics' coefficients (with the statistical significance). $Constant$ (i.e., $\theta_0$) is the expected value of \textit{Recall@20}.}
\label{tab:rq1}
\begin{adjustbox}{width=\textwidth,center}
    \centering
    \begin{tabular}{l|cc|cc|cc|cc}
    \toprule
        \multicolumn{1}{c}{\multirow{2}{*}{\textbf{Characteristics}}} & \multicolumn{2}{c}{\textbf{LightGCN}} & \multicolumn{2}{c}{\textbf{DGCF}} & \multicolumn{2}{c}{\textbf{UltraGCN}} & \multicolumn{2}{c}{\textbf{SVD-GCN}} \\ \cmidrule{2-9} \multicolumn{1}{c}{} & \multicolumn{1}{c}{\textbf{Yelp2018}} & \multicolumn{1}{c}{\textbf{Gowalla}} & \textbf{Yelp2018} & \multicolumn{1}{c}{\textbf{Gowalla}} & \textbf{Yelp2018} & \multicolumn{1}{c}{\textbf{Gowalla}} & \textbf{Yelp2018} & \textbf{Gowalla} \\ \cmidrule{1-9}
        \rowcolor{gray} \ $R^{2}$(adj. $R^{2}$) & $0.971 (0.971)$& $0.979 (0.978)$& $0.973 (0.973)$& $0.982 (0.981)$& $0.965 (0.964)$& $0.860 (0.858)$& $0.982 (0.981)$& $0.981 (0.981)$
 \\ \
         $Constant$ & $0.100^{***}$&$0.121^{***}$&$0.089^{***}$&$0.107^{***}$&$0.061^{***}$&$0.062^{***}$&$0.116^{***}$&$0.135^{***}$
 \\ \
         $SpaceSize_{log}$ & $0.070^{***}$& $0.192^{***}$& $0.133^{***}$& $0.237^{***}$& $-0.059^{***}$& $0.318^{***}$& $0.064^{***}$& $0.114^{***}$
 \\ \
         $Shape_{log}$ & $-0.253^{*}$&$-0.231^{}$&$-0.282^{**}$&$-0.220^{*}$&$0.135^{}$&$-0.003^{}$&$-0.193^{}$&$-0.232^{*}$
 \\ \
         $Density_{log}$ & $0.194^{***}$&$0.298^{***}$& $0.243^{***}$& $0.327^{***}$& $0.026^{*}$& $0.321^{***}$& $0.203^{***}$& $0.234^{***}$
 \\ \
         $Gini\text{-}U$ & $0.296^{**}$&$0.104^{}$&$0.074^{}$&$-0.071^{}$&$-0.043^{}$&$-0.931^{***}$&$0.136^{}$&$0.143^{}$
 \\ \
         $Gini\text{-}I$ & $1.362^{***}$& $0.681^{***}$& $1.108^{***}$& $0.560^{***}$& $0.605^{***}$& $-0.144^{}$& $1.138^{***}$& $0.748^{***}$
 \\ \
         $AvgDegree\text{-}U_{log}$ & $0.390^{***}$&$0.605^{***}$&$0.518^{***}$&$0.673^{***}$&$-0.100^{*}$&$0.640^{***}$&$0.364^{***}$&$0.464^{***}$
 \\ \
         $AvgDegree\text{-}I_{log}$ & $0.137^{*}$& $0.374^{***}$& $0.235^{***}$& $0.453^{***}$& $0.034^{}$& $0.637^{***}$& $0.171^{**}$& $0.231^{**}$
 \\ \
         $AvgClustC\text{-}U_{log}$ & $0.613^{***}$&$0.665^{***}$&$0.726^{***}$&$0.783^{***}$&$-0.077^{}$&$0.706^{***}$&$0.613^{***}$&$0.496^{***}$
 \\ \
         $AvgClustC\text{-}I_{log}$ & $0.087^{}$& $0.332^{*}$& $0.168^{}$& $0.373^{**}$& $0.062^{}$& $0.671^{**}$& $0.057^{}$& $0.215^{}$
 \\ \
        $Assort\text{-}U$ & $0.094^{***}$&$0.024^{*}$&$0.093^{***}$&$0.013^{}$&$0.123^{***}$&$-0.019^{}$&$0.080^{***}$&$0.010^{}$
 \\ \
         $Assort\text{-}I$ & $-0.051^{}$& $-0.031^{}$& $-0.056^{*}$& $-0.055^{}$& $0.001^{}$& $-0.174^{***}$& $-0.048^{*}$& $-0.088^{*}$ \\ 
   \bottomrule
 \multicolumn{9}{l}{\textit{***p-value $\leq$ 0.001, **p-value $\leq$ 0.01, *p-value $\leq$ 0.05}}
    \end{tabular}
    \end{adjustbox}
\end{table}

\noindent \textbf{Topological dataset characteristics.} 
Graph-based recommendation systems interpret the user-item interaction data as a bipartite and undirected graph. Consequently, this study assesses the influence of \textit{topological} characteristics of the selected recommendation datasets on the accuracy performance.

The most evident outcome is that $AvgDegree\text{-}U_{log}$ and $AvgDe\-gree\text{-}I_{log}$ show a direct correspondence with recommendation accuracy performance in almost all settings. Indeed, this analytically confirms what we already observed in Section \ref{sec:top-char-graph-collab} regarding the explicit presence of the node degree in the formulations of all the selected graph-based approaches. In practical terms, when graph-based models are trained on datasets with several interactions for users and items, they learn accurate users' preferences since each node receives the contribution of numerous neighbor nodes. 
It is worth noticing that, in absolute values, $AvgDegree\text{-}U_{log}$ is more influential than $AvgDegree\text{-}I_{log}$ on the overall performance. 
Hence, under the same average degree gain, an improvement in the user average degree is preferable since it would more significantly improve the overall performance.

As far as clustering coefficient and degree assortativity are concerned, we assess how similarities among nodes from the same partition in the graph may impact the recommendation accuracy performance of models. In terms of $AvgClustC\text{-}U_{log}$ and $AvgClustC\text{-}I_{log}$, the results prove again a strong direct correspondence in almost all settings of graph-based models and datasets. Differently from the average degree scenario, the relative importance of the user-side values is much higher than the one of the item-side for LightGCN and DGCF, while the gap sometimes gets narrower in the case of UltraGCN and SVD-GCN. This may be because while LightGCN and DGCF only leverage user-item types of interactions, UltraGCN and (especially) SVD-GCN also embed the information conveyed in the user- and item-projected graphs in their formulations, thus flattening the different influence of the user-side characteristics over the item-side counterpart. 

Interestingly, the $Assort\text{-}U$ and $Assort\text{-}I$ characteristics exhibit a direct and inverse correspondence to the accuracy metric, respectively. Furthermore, models such as LightGCN and DGCF have slightly larger coefficients for both $Assort\text{-}U$ and $Assort\text{-}I$ than SVD-GCN. Again, these results have a mathematical justification. Indeed, the strong \textit{lookahead} nature of the assortativity measures (refer again to Section \ref{sec:assortativity}) seems to be captured by the multi-layer message-passing performed by LightGCN and DGCF.
Conversely, in the case of SVD-GCN, they are less influential, probably because the model acts on the singular values of the adjacency matrix with the effect of limiting the graph convolutional layers' depth to avoid over-smoothing.
However, it is important to observe that the assortativity results are less statistically-significant than the others, so we plan to further investigate this aspect in future work. 
On the contrary, a different trend can be observed for UltraGCN, where both $Assort\text{-}U$ and $Assort\text{-}I$ generally have bigger coefficients with much more statistically significant values. Again, this behavior could have a theoretical foundation since the model adopts the infinite-layer approximation, which (differently from SVD-GCN) may capture long-distance relationships in the user-item graph.

\noindent\textsc{\bfseries Summary.} \textit{The analytical and theoretical observations show that: (i) factorization-based approaches could be the core component of graph-based recommender systems; (ii) while confirming its influence on the recommendation performance, node degree seems not to be a key topological characteristic to distinguish among the different graph-based models; indeed, the wider perspective provided by clustering coefficient and (especially) degree assortativity may help to recognize how the different models address the topological properties of the graph, even with unexpected outcomes.}

\subsection{Influence of node- and edge-dropout (RQ2)} 
\label{sec:rq2}
The current section investigates the influence of the different sampling strategies, namely, node- and edge-dropout, on the explanatory model. Given the lack of space, we report an extensive analysis of the largest dataset, Gowalla, by considering the performance of LightGCN and SVD-GCN.

As generally observed in real-world networks, user-item bipartite graphs follow the typical trend of \textit{scale-free} networks~\cite{PhysRevE.67.026112}. Thus, we introduce the following statement on the possible influence of node- and edge-dropout on our linear explanatory model:

\noindent\textsc{\bfseries Statement.} \textit{
In general, the node-dropout strategy drops wider portions of the original topology than the edge-dropout, so it may negatively impact the significance of the linear model explanations.}

The interested reader may refer to Appendix \ref{app:intuition} for the empirical intuition which drove us to the reported statement. To analytically test the statement, we build four versions of the dataset $\mathbf{X}_c$, each with varying portions of sub-datasets generated through node and edge dropout, respectively (refer to Equation \ref{base_formula}). Specifically, the number of samples in $\mathbf{X}_c$ changes in accordance to:
\begin{equation}
    |\mathbf{X}_c| = (1 - \alpha) |\mathbf{X}^{n}_{c}| + \alpha |\mathbf{X}^{e}_{c}|,
\end{equation}
where $\mathbf{X}^{n}_{c}$ and $\mathbf{X}^{e}_{c}$ indicate the portion of $\mathbf{X}_c$ sampled through node- and edge-dropout, while $\alpha$ is a parameter to control the number of samples from $\mathbf{X}^{n}_{c}$ and $\mathbf{X}^{e}_{c}$ contributing to the final dataset $\mathbf{X}_c$. We use $|\cdot|$ as a necessary notation abuse to refer to any dataset size in a simple way. We let $\alpha$ range in $\{0.0, 0.3, 0.7, 1.0\}$, where extreme values of $\alpha$ are used to build the dataset through either node- or edge-dropout; the others combine the two sampling strategies.

Table \ref{tab:rq2} reports the explanatory model results for the \textit{Recall@20} as accuracy metric at varying $\alpha$ values, along with the average sampling statistics on each setting of $\alpha$. The reader may refer to Appendix \ref{app:add_exp_2} for additional experiments reporting the performance of DGCF and UltraGCN on Gowalla. In alignment with the above statement, the average sampling statistics show that node-dropout generally retains smaller portions of the graph than the edge-dropout. Then, the regression results highlight that the optimal trade-off between high $R^2$ (adj. $R^2$) and statistical significance of the learned coefficients is reached when combining samples generated through both node- and edge-dropout. On the contrary, the settings with either node- or edge-dropout do not offer the conditions for the regression model to learn meaningful dependencies, with respect to the $R^2$ (adj. $R^2$) and/or statistical significance. Indeed, in the extreme cases, the characteristic-performance dependencies are not aligned with the ones observed in RQ1 either on the sign or on the absolute value of the coefficients. 
This justifies the dataset sampling adopted to explore RQ1.

\noindent\textsc{\bfseries Summary.} \textit{The empirical and analytical evaluation of the explanatory model for different settings of node- and edge-dropout indicates that their simultaneous combination to generate the sub-datasets (i.e., the strategy we followed in RQ1) is beneficial to produce meaningful explanations.}

\begin{table}[!t]
\caption{
Results of the explanatory model on Gowalla (\textit{Recall@20}) obtained with LightGCN and SVD-GCN, for different proportions of sub-datasets generated through node- and edge-dropout. The header reports a graphical intuition of $\alpha$'s variation and average sampling statistics.
}
\label{tab:rq2}
\begin{adjustbox}{width=\textwidth,center}
    \centering
    \begin{tabular}{l|cc|cc|cc|cc}
    \toprule
         \multicolumn{1}{c}{} & \multicolumn{2}{c}{\textbf{Node drop} \tikzcircle[fill=black]{2.5pt} \tikzcircle[fill=black]{2.5pt} \tikzcircle[fill=black]{2.5pt} \quad
         \textbf{Edge drop} \tikzcircle[fill=white]{2.5pt} \tikzcircle[fill=white]{2.5pt} \tikzcircle[fill=white]{2.5pt}} & \multicolumn{2}{c}{\textbf{Node drop} \tikzcircle[fill=black]{2.5pt} \tikzcircle[fill=black]{2.5pt} \tikzcircle[fill=white]{2.5pt} \quad
         \textbf{Edge drop} \tikzcircle[fill=black]{2.5pt} \tikzcircle[fill=white]{2.5pt} \tikzcircle[fill=white]{2.5pt}} & \multicolumn{2}{c}{\textbf{Node drop} \tikzcircle[fill=black]{2.5pt} \tikzcircle[fill=white]{2.5pt} \tikzcircle[fill=white]{2.5pt} \quad
         \textbf{Edge drop} \tikzcircle[fill=black]{2.5pt} \tikzcircle[fill=black]{2.5pt} \tikzcircle[fill=white]{2.5pt}} & \multicolumn{2}{c}{\textbf{Node drop} \tikzcircle[fill=white]{2.5pt} \tikzcircle[fill=white]{2.5pt} \tikzcircle[fill=white]{2.5pt} \quad
         \textbf{Edge drop} \tikzcircle[fill=black]{2.5pt} \tikzcircle[fill=black]{2.5pt} \tikzcircle[fill=black]{2.5pt}} \\ \cmidrule{2-9}
         \multicolumn{1}{c}{} & \multicolumn{2}{c}{\makecell[c]{\textit{\ul{Average Sampling Statistics}}\\\textbf{Users:} 5,828 \quad \textbf{Items:} 7,887 \\ \textbf{Interactions:} 45,620}} & \multicolumn{2}{c}{\makecell[c]{\textit{\ul{Average Sampling Statistics}}\\\textbf{Users:} 12,744 \quad \textbf{Items:} 17,229 \\ \textbf{Interactions:} 97,785}} & \multicolumn{2}{c}{\makecell[c]{\textit{\ul{Average Sampling Statistics}}\\\textbf{Users:} 21,730 \quad \textbf{Items:} 29,316 \\ \textbf{Interactions:} 160,919\\}} & \multicolumn{2}{c}{\makecell[c]{\textit{\ul{Average Sampling Statistics}}\\\textbf{Users:} 28,526 \quad \textbf{Items:} 38,467 \\ \textbf{Interactions:} 209,659}} \\ \cmidrule{2-9} 
         \multicolumn{1}{c}{\textbf{Characteristics}} & \multicolumn{1}{c}{\textbf{LightGCN}} & \multicolumn{1}{c}{\textbf{SVD-GCN}} & \multicolumn{1}{c}{\textbf{LightGCN}} & \multicolumn{1}{c}{\textbf{SVD-GCN}} & \multicolumn{1}{c}{\textbf{LightGCN}} & \multicolumn{1}{c}{\textbf{SVD-GCN}} & \multicolumn{1}{c}{\textbf{LightGCN}} & \multicolumn{1}{c}{\textbf{SVD-GCN}} \\ 
         \cmidrule{1-9} 
         \rowcolor{gray} $R^{2}$(adj. $R^{2}$) &$0.597 (0.583)$&$0.754 (0.745)$ & $0.968 (0.967)$&$0.970 (0.969)$ & $0.986 (0.985)$&$0.987 (0.987)$ & $0.994 (0.994)$&$0.991 (0.991)$ \\
        $Constant$ & $0.179^{***}$&$0.193^{***}$ & $0.146^{***}$&$0.159^{***}$ & $0.098^{***}$&$0.112^{***}$ & $0.062^{***}$&$0.077^{***}$ \\
        $SpaceSize_{log}$ & $0.092^{}$&$0.037^{}$ & $0.143^{***}$&$0.064^{**}$ & $0.220^{***}$&$0.136^{***}$ & $-0.162^{***}$&$-0.048^{}$ \\
        $Shape_{log}$ & $-0.078^{}$&$-0.118^{}$ & $-0.265^{}$&$-0.261^{*}$ & $-0.175^{}$&$-0.303^{}$ & $0.084^{}$&$-0.157^{}$ \\
        $Density_{log}$ & $-0.079^{**}$&$-0.090^{**}$ & $0.261^{***}$&$0.195^{***}$ & $0.323^{***}$&$0.251^{***}$ & $0.072^{**}$&$0.040^{}$ \\
        $Gini\text{-}U$ & $-0.013^{}$&$0.015^{}$ & $0.184^{}$&$0.193^{}$ & $0.246^{}$&$0.224^{}$ & $0.291^{***}$&$0.225^{*}$ \\
        $Gini\text{-}I$ & $0.883^{***}$&$0.884^{***}$ & $0.856^{***}$&$0.867^{***}$ & $0.911^{***}$&$0.940^{***}$ & $0.389^{***}$&$0.387^{***}$ \\
        $AvgDegree\text{-}U_{log}$ & $0.052^{}$&$0.005^{}$ &$0.536^{***}$&$0.390^{***}$ & $0.631^{***}$&$0.539^{***}$ & $-0.132^{*}$&$0.070^{}$ \\
        $AvgDegree\text{-}I_{log}$ & $-0.026^{}$&$-0.112^{}$ &$0.271^{**}$&$0.129^{}$ & $0.456^{***}$&$0.236^{*}$ & $-0.048^{}$&$-0.087^{}$ \\
        $AvgClustC\text{-}U_{log}$ & $0.209^{}$&$0.168^{}$ & $0.654^{***}$&$0.508^{**}$ & $0.687^{**}$&$0.647^{***}$ & $-0.133^{}$&$0.016^{}$ \\
        $AvgClustC\text{-}I_{log}$ & $-0.141^{}$&$-0.227^{}$ & $0.137^{}$&$-0.007^{}$ & $0.436^{}$&$0.172^{}$ & $-0.145^{}$&$-0.112^{}$ \\
        $Assort\text{-}U$ & $0.008^{}$&$-0.001^{}$ & $0.017^{}$&$0.008^{}$ & $0.013^{}$&$-0.002^{}$ & $0.011^{}$&$-0.003^{}$ \\
        $Assort\text{-}I$ & $-0.022^{}$&$-0.078^{**}$ & $0.028^{}$&$-0.057^{}$ & $0.059^{}$&$-0.056^{}$ & $0.012^{}$&$-0.037^{}$
         \\
    \bottomrule
    \multicolumn{9}{l}{\textit{***p-value $\leq$ 0.001, **p-value $\leq$ 0.01, *p-value $\leq$ 0.05}}
    \end{tabular}
    \end{adjustbox}
\end{table}
%\section{Conclusion and Future Work}
\section{Future directions}
%We plan to extend the proposed analysis to assess the impact of topological dataset characteristics on other recommendation metrics accounting for the novelty and diversity of the produced recommendation lists, and potential biases and fairness issues in recommendation; this investigation may be conducted by utilizing ad-hoc graph recommender systems specifically designed to optimize such objectives. Furthermore, given the interesting insights from RQ2, we intend to directly create the synthetic user-item graphs from scratch with graph generator techniques which may resemble real-world recommendation data but with desired topological properties. Finally, we seek to investigate the impact of other topological aspects of the user-item graph on the performance of graph-based recommender systems, such as the presence of user, item, and user-item communities and their inter dependencies, as this direction has been poorly explored in the related literature so far.

We plan to assess the impact of topological dataset characteristics on beyond-accuracy metrics (i.e., novelty and diversity of the produced recommendation lists, and potential biases and fairness issues in recommendation). Furthermore, given the interesting insights from RQ2, we intend to directly create synthetic user-item graphs from scratch with graph generator techniques which may resemble real-world recommendation data but with desired topological properties. In addition to this, we seek to investigate the impact of other topological aspects of the user-item graph on the performance of graph-based recommender systems (such as the presence of communities and their inter dependencies) as well as explore other sampling strategies based, for instance, on a temporal time-split. Finally, we plan to formalize and define guidelines to design and implement novel graph-based recommender systems which exploit the lessons-learned from the conducted analysis.

\bibliographystyle{unsrtnat}
\bibliography{reference}

\appendix
\section{Appendix}

\subsection{Classical dataset characteristics}
\label{app:class_charact}

We provide a brief overview of popular classical characteristics describing a recommendation dataset (as presented in~\cite{DBLP:journals/tmis/AdomaviciusZ12, DBLP:conf/sigir/DeldjooNSM20}). Regarding the adopted notation, we assume the reader is aware of the background from Section \ref{sec:topological-characteristics}.

\subsubsection{Space size} The space size estimates the number of all possible interactions that might exist among users and items:

\begin{equation}
    \zeta = \sqrt{UI}.
\end{equation}

\subsubsection{Shape} The shape of a recommendation dataset is defined as the ratio between the number of users and items:

\begin{equation}
    \pi = \frac{U}{I}.
\end{equation}

\subsubsection{Density} The density of a recommendation dataset measures the ratio of actual user-item interactions with respect to all possible interactions that might connect all users and items:

\begin{equation}
    \delta = \frac{E}{UI},
\end{equation}
where $E = |\{(u, i)\;|\;\mathbf{R}_{u, i} = 1\}|$ is the number of interactions existing among users and items in the recommendation data.

\subsubsection{Gini coefficient} 
\label{sec:gini}
The Gini coefficient is an estimation of the interactions' concentration for both users and items. When calculated on $\mathcal{U}$ and $\mathcal{I}$, we have:
\begin{equation}
    \kappa_{\mathcal{U}} = \frac{\sum\limits_{u = 1}^{U-1}\sum\limits_{v = u + 1}^{U}abs(\sigma_u - \sigma_v)}{U\sum\limits_{u = 1}^{U}\sigma_u},
    \qquad \kappa_{\mathcal{I}} = \frac{\sum\limits_{i = 1}^{I-1}\sum\limits_{j = i + 1}^{I}abs(\sigma_i - \sigma_j)}{I\sum\limits_{i = 1}^{I}\sigma_i}, 
\end{equation}
where $abs()$ is the function returning the absolute value. 

\subsection{(Extended) Topological characteristics in graph collaborative filtering}
\label{app:graph-recsys}
In this section, we present the four selected graph-based recommender systems used in our analysis, and re-formulate their techniques to make the \textit{topological} data characteristics explicitly emerge. As additional background with respect to Section \ref{sec:topological-characteristics}, we introduce the notations $\mathbf{e}_u \in \mathbb{R}^b$ and $\mathbf{e}_i \in \mathbb{R}^b$ as the initial embeddings of the nodes for user $u$ and item $i$, respectively, where $b << U, I$. Then, in the case of message-propagation at different layers, we also introduce the notations $\mathbf{e}_u^{(l)}$ and $\mathbf{e}_i^{(l)}$ to indicate the updated node embeddings for user $u$ and item $i$ after $l$ propagation layers, with $0 \leq l \leq L$ (note that $\mathbf{e}_u^{(0)} = \mathbf{e}_u$ and $\mathbf{e}_i^{(0)} = \mathbf{e}_i$).

\subsubsection{LightGCN} \citet{DBLP:conf/sigir/0001DWLZ020} propose to lighten the graph convolutional layer presented in~\citet{DBLP:conf/iclr/KipfW17} for the recommendation task. Specifically, their layer removes feature transformation and non-linearities:
\begin{equation}
\label{eq:lightgcn}
    \mathbf{e}_u^{(l)} = \sum\limits_{i' \in \mathcal{N}_u^{(1)}} \frac{A_{ui'} \mathbf{e}_{i'}^{(l - 1)}}{\sqrt{\sigma_u  \sigma_{i'}}}, \quad \mathbf{e}_i^{(l)} = \sum\limits_{u' \in \mathcal{N}_i^{(1)}} \frac{A_{iu'} \mathbf{e}_{u'}^{(l - 1)}}{\sqrt{\sigma_i \sigma_{u'}}},
\end{equation}
where each neighbor contribution is weighted through the corresponding entry in the normalized Laplacian adjacency matrix to flatten the differences among nodes with high and low degrees. Since $A_{ui'} = 1, \text{ } \forall i' \in \mathcal{N}^{(1)}_u$ (the dual holds for $A_{iu'}$), the contribution weighting comes only from the denominator.  

\subsubsection{DGCF} \citet{DBLP:conf/sigir/WangJZ0XC20} assume that user-item interactions are decomposed into a set of independent intents, representing the specific aspects users may be interested in when interacting with items. In this respect, the authors propose to iteratively learn a set of weighted adjacency matrices $\{\mathbf{\tilde{A}}_1, \mathbf{\tilde{A}}_2, \dots\}$, where each of them records the user-item importance weights based on the specific intent it represents. Then, they introduce a graph disentangling layer for each weighted adjacency matrix:
\begin{equation}
    \mathbf{e}_{u, *}^{(l)} = \sum\limits_{i' \in \mathcal{N}_u^{(1)}} \frac{\tilde{A}_{ui', *} \mathbf{e}_{i', *}^{(l - 1)}}{\sqrt{\sigma_{u, *} \sigma_{i', *}}}, \quad \mathbf{e}_{i, *}^{(l)} = \sum\limits_{u' \in \mathcal{N}_i^{(1)}} \frac{\tilde{A}_{iu', *} \mathbf{e}_{u', *}^{(l - 1)}}{\sqrt{\sigma_{i, *} \sigma_{u', *}}},
\end{equation}
where $\tilde{A}_{ui', *}$ and $\mathbf{e}_{i',*}^{(l-1)}$ are the learned importance weight of user $u$ on item $i'$ and the embedding of item $i'$ for any intent, while $\sigma_{u,*}$ is the corresponding node degree calculated on $\mathbf{\tilde{A}_{*}}$ (the same applies for the item side).    

\subsubsection{UltraGCN} \citet{DBLP:conf/cikm/MaoZXLWH21} recognize three major limitations in GCN-based message-passing for collaborative filtering, namely, (i) the asymmetric weight assignment to connected nodes when considering user-user and item-item relationships; (ii) the impossibility to diversify the importance of each type of relation (i.e., user-item, user-user, item-item) during the message-passing; (iii) the over-smoothing effect when stacking more than 3 layers. To tackle such issues, the authors propose to go beyond the traditional concept of explicit message-passing, and approximate the infinite-layer message-passing through the following:
\begin{equation}
    \mathbf{e}_u = \sum\limits_{i' \in \mathcal{N}_u^{(1)}} \frac{A_{ui'} \sqrt{\sigma_u + 1} \mathbf{e}_{i'}}{\sigma_u \sqrt{\sigma_{i'} + 1}}, \quad
    \mathbf{e}_i = \sum\limits_{u' \in \mathcal{N}_i^{(1)}} \frac{A_{iu'} \sqrt{\sigma_i + 1} \mathbf{e}_{u'}}{\sigma_{i} \sqrt{\sigma_{u'} + 1}}.
\end{equation}
Note that the procedure is not repeated for layers $l > 1$, as the method surpasses the concept of iterative message-passing. During the optimization, the model first minimizes a constraint loss that adopts negative sampling to limit the over-smoothing effect:
\begin{equation}
    \begin{aligned}
    \min\limits_{\mathbf{e}_u, \mathbf{e}_i, \mathbf{e}_j} \text{ }&-  \sum\limits_{(u,i) \in \mathbf{R}_s^{+}}\frac{1}{\sigma_u} \frac{\sqrt{\sigma_u + 1}}{\sqrt{\sigma_i + 1}} log(sig(\mathbf{e}_u^\top \cdot \mathbf{e}_i)) \text{ }+ \\
    & - \sum\limits_{(u, j) \in \mathbf{R}_s^{-}} \frac{1}{\sigma_u} \frac{\sqrt{\sigma_u + 1}}{\sqrt{\sigma_j + 1}} log(sig(-\mathbf{e}_u^\top \cdot \mathbf{e}_j)),
    \end{aligned}
\end{equation}
where $\mathbf{R}_s^{+}$ and $\mathbf{R}^{-}_s$ are pairs of positive and negative interactions sampled from the user-item matrix $\mathbf{R}$, while $log()$ and $sig()$ are the logarithm and sigmoid function. Then, they also take into account the item-projected graph $\mathcal{G}_{\mathcal{I}}$ and minimize the following:
\begin{equation}
    \min\limits_{\mathbf{e}_u, \mathbf{e}_j} \text{ }- \sum\limits_{(u, i) \in \mathbf{R}^{+}_s} \text{ } \sum\limits_{j \in topk(\mathbf{R}^{\mathcal{I}}_{i, *})}\frac{\mathbf{R}^{\mathcal{I}}_{i, j}}{\sigma^{\mathcal{I}}_i - \mathbf{R}^{\mathcal{I}}_{i,i}} \sqrt{\frac{\sigma^{\mathcal{I}}_i}{\sigma^{\mathcal{I}}_j}} log(sig(\mathbf{e}_u^\top \cdot \mathbf{e}_j)),
\end{equation}
where $topk()$ retrieves the top-k values of a matrix row-wise, and $\sigma_*^{\mathcal{I}}$ is the node degree calculated on $\mathcal{G}_{\mathcal{I}}$. 

\subsubsection{SVD-GCN} \citet{DBLP:conf/cikm/PengSM22} propose a reformulation of the GCN-based message-passing which leverages the similarities between graph convolutional layers and singular value decomposition (i.e., SVD). Specifically, they rewrite the message-passing introduced in LightGCN by making two aspects explicitly emerge, namely: (i) even- and odd-connection message aggregations, and (ii) singular values and vectors obtained by decomposing the user-item interaction matrix $\mathbf{R}$ through SVD. On such basis, the authors' assumption is that the traditional graph convolutional layer intrinsically learns a low-rank representation of the user-item interaction matrix where components corresponding to larger singular values tend to be enhanced. %Moreover, t
They reinterpret the over-smoothing effect as an increasing gap between singular values when stacking more and more layers. 
%In the light of this, 
The embeddings for users and items are obtained as follows:
\begin{equation}
    \mathbf{e}_u = \mathbf{p}_u exp(a_1 \boldsymbol\lambda) \cdot \mathbf{W}, \qquad
    \mathbf{e}_i = \mathbf{q}_i exp(a_1 \boldsymbol\lambda) \cdot \mathbf{W},
\end{equation}
where: (i) $\mathbf{p}_u$ and $\mathbf{q}_i$ are the left and right singular vectors of the normalized user-item interaction matrix for user $u$ and item $i$; (ii) $exp()$ is the exponential function; (iii) $a_1$ is a tunable hyper-parameter of the model; (iv) $\boldsymbol\lambda$ is the vector of the largest singular values of the normalized user-item matrix; (v) $\mathbf{W}$ is a trainable matrix to perform feature transformation. Note that the highest singular value $\lambda_{max}$ and the maximum node degree $\max(\mathcal{D})$ in the user-item interaction matrix are associated by the following inequality:
\begin{equation}
    \lambda_{max} \leq \frac{\max(\mathcal{D})}{\max(\mathcal{D}) + a_2},
\end{equation}
where $a_2$ is another tunable hyper-parameter of the model to control the gap among singular values. Similarly to UltraGCN, the authors recognize the importance of different types of relationships during the message-passing (i.e., user-item, user-user, item-item). For this reason, they decide to augment the loss function with other components addressing also the similarities among node embeddings from the same partition:
\begin{equation}
    \begin{aligned}
        \min\limits_{\mathbf{e}_v, \mathbf{e}_w, \mathbf{e}_j} \text{ }&- \sum\limits_{(v, w) \in (\mathbf{R}^{\mathcal{*}}_s)^{+}} log(sig(\mathbf{e}^\top_v \cdot \mathbf{e}_w)) \text{ }+ \\
        &- \sum\limits_{(v, j) \in (\mathbf{R}^{\mathcal{*}}_s)^{-}} log(sig(-\mathbf{e}^\top_v \cdot \mathbf{e}_j)),
    \end{aligned}
\end{equation}
where $v$, $w$, and $j$ are nodes from the same partition, and $\mathbf{R}^*$ is the interaction matrix of that partition.

\subsection{Algorithms}
\label{app:alg}
In Algorithm \ref{alg:sampling} and Algorithm \ref{alg:data-generation} we report the pseudocode to perform graph sampling and the sub-dataset generation, respectively. Note that $uniform_N()$ is the function that uniformly samples $N$ elements from a set, while $uniform()$ samples only one element from a set.

\begin{algorithm}[!h]
\caption{Graph sampling}
\label{alg:sampling}
\SetAlgoLined
\textbf{Input:} Bipartite user-item graph $\mathcal{G}$, dropout rate $\mu$, graph sampling strategy \texttt{\textit{sampling}}.\\
\textbf{Output:} Sampled graph $\mathcal{G}_m$.\\
\uIf{\texttt{sampling} $==$ \texttt{nodeDropout}}{
    %\texttt{// calculate nodes to sample according to} $d$\\
    $N = (U + I) * (1 - \mu)$ \\
    %\texttt{// sample $N$ nodes}\\
    $\mathcal{V}_m \leftarrow uniform_N(\mathcal{U} \cup \mathcal{I})$\\
    %\texttt{// mask $\mathbf{A}$ according to $\mathcal{V}_m$}\\
    $\mathbf{A}_m \leftarrow mask_{node}(\mathbf{A}, \mathcal{V}_m)$
}
\uElseIf{\texttt{sampling} $==$ \texttt{edgeDropout}}{
    %\texttt{// calculate edges to sample according to} $d$\\
    $N = E * (1 - \mu)$ \\
    %\texttt{// sample $N$ edges}\\
    $\mathcal{E}_m \leftarrow uniform_N(\mathcal{E}_{u \rightarrow i})$\\
    %\texttt{// mask $\mathbf{A}$ according to $\mathcal{E}_m$}\\
    $\mathbf{A}_m \leftarrow mask_{edge}(\mathbf{A}, \mathcal{E}_m)$\\
    %\texttt{// induce the set of nodes $\mathcal{V}_m$}\\
    $\mathcal{V}_m \leftarrow induce(\mathbf{A}_m)$
}
\phantom{ }\\
%\texttt{// induce $\mathcal{G}_m$ through $\mathcal{V}_m$ and $\mathbf{A}_m$}\\
$\mathcal{G}_m \leftarrow \{\mathcal{V}_m, \mathbf{A}_m\}$ \\
Return $\mathcal{G}_m$.
\end{algorithm}
\begin{algorithm}[!h]
\SetAlgoLined
\textbf{Input:} Bipartite user-item graph $\mathcal{G}$, number of samples $M$.\\
\textbf{Output:} $M$ sampled graphs.\\
$m \leftarrow 1$\\
$\mathcal{M} = \{\}$\\
\While{$m \leq M$}{
    % \texttt{// get dropout rate}\\
    $\mu \leftarrow uniform([0.7, 0.9])$\\
    % \texttt{// choose graph sampling strategy}\\
    $\texttt{\textit{sampling}} \leftarrow uniform(\{\texttt{nodeDropout}, \texttt{edgeDropout}\})$\\
    % \texttt{// add sampled graph to the set}\\
    $\mathcal{M} \leftarrow \mathcal{M} \cup sample(\mathcal{G}, \mu, \texttt{\textit{sampling}})$\\
    % \texttt{// increment number of generated samples}\\
    $m \leftarrow m + 1$
}
Return $\mathcal{M}$.
\caption{Sub-dataset generation.}
\label{alg:data-generation}
\end{algorithm}

\subsection{Experimental setting}
\label{app:exp_sett}

We provide here a detailed description of the experimental settings for our proposed explanatory framework. First, we present the recommendation datasets for this study. Then, we report on the adopted characteristics, along with details about their (optional) value rescaling and denomination. Finally, we describe the methodology we follow to train and evaluate the graph-based recommendation models to foster the reproducibility of this work.

\subsubsection{Recommendation datasets} We use specific versions of Yelp2018~\cite{DBLP:conf/cikm/PengSM22} and Gowalla~\cite{DBLP:conf/sigir/0001DWLZ020} (whose results are reported in the main paper), plus Amazon-Book~\cite{DBLP:conf/www/WangHWYL0C21} (whose results are only reported in this Appendix to stay within the page limits). The usage of such datasets is motivated by their popularity in graph collaborative filtering~\cite{DBLP:conf/sigir/0001DWLZ020, DBLP:conf/cikm/MaoZXLWH21, DBLP:conf/www/LinTHZ22, DBLP:conf/cikm/GongSWLL22}. Yelp2018~\cite{DBLP:journals/corr/Asghar16} collects data about users and businesses interactions, Amazon-Book is a sub-category of the Amazon dataset~\cite{DBLP:conf/www/HeM16}, and Gowalla~\cite{DBLP:conf/kdd/ChoML11} is a social-based dataset where users share their locations. Note that, to provide a coherent calculation of the characteristics, we retained only the subset of nodes and edges for each dataset which induces the widest connected graph. In the following, we present the calculation of characteristics, experimentally justifying their adoption.

\subsubsection{Characteristics calculation}
\label{app:char-calc}
Following the same setting as in~\cite{DBLP:conf/sigir/DeldjooNSM20}, we generate $M = 600$ sub-datasets from the original ones through the techniques described in Algorithm \ref{alg:sampling} and Algorithm ~\ref{alg:data-generation}, resulting in a total of 1,800 synthetic samples (if we consider all the three datasets in the main paper and in the Appendix). Second, inspired by similar works~\cite{DBLP:journals/tmis/AdomaviciusZ12, DBLP:conf/sigir/DeldjooNSM20}, we decide to apply the log10-scale to the formulation of some characteristics to obtain values within comparable order of magnitude, thus making the training of the explanatory model more stable. In Table \ref{tab:shorthand} we provide a comprehensive outlook on the set of characteristics, where we apply a renaming scheme for the sake of simple understanding and reference. Furthermore, Table \ref{tab:dataset-chars} displays the statistics of the overall datasets and the aggregated characteristics for the generated samples. Finally, Figure \ref{fig:correlation} empirically supports the usage of the selected characteristics, as they appear loosely correlated.

\begin{table}[!b]
\caption{Selected \textit{classical} and \textit{topological} characteristics. We report the full name, the symbol, whether it is rescaled via log10, and the shorthand adopted.}
\label{tab:shorthand}
\begin{adjustbox}{width=0.8\textwidth,center}
\begin{tabular}{l|l|c|c|l}
\toprule
\multicolumn{1}{l}{\textbf{Type}} & \multicolumn{1}{l}{\textbf{Characteristics}} & \multicolumn{1}{l}{\textbf{Symbol}} & \multicolumn{1}{l}{\textbf{Log10}} & \multicolumn{1}{l}{\textbf{Shorthand}} \\
 \cmidrule{1-5}
\multirow{5}{*}{\textit{Classical}} & {Space size} & {$\zeta$} & {\checkmark} & {$SpaceSize_{log}$} \\
 & Shape & $\pi$ & \checkmark & $Shape_{log}$ \\
 & {Density} & {$\delta$} & {\checkmark} & {$Density_{log}$} \\
 & Gini user & $\kappa_{\mathcal{U}}$ &  & $Gini\text{-}U$ \\
 & {Gini item} & {$\kappa_{\mathcal{I}}$} & {} & {$Gini\text{-}I$} \\
 \hline
\multirow{6}{*}{\textit{Topological}} & {Average degree user} & {$\sigma_{\mathcal{U}}$} & {\checkmark} & {$AvgDegree\text{-}U_{log}$} \\
 & Average degree item & $\sigma_{\mathcal{I}}$ & \checkmark & $AvgDegree\text{-}I_{log}$ \\
 & {Average clustering coefficient user} & {$\gamma_{\mathcal{U}}$} & {\checkmark} & {$AvgClustC\text{-}U_{log}$} \\
 & Average clustering coefficient item & $\gamma_{\mathcal{I}}$ & \checkmark & $AvgClustC\text{-}I_{log}$ \\
 & {Degree assortativity user} & {$\rho_{\mathcal{U}}$} & {} & {$Assort\text{-}U$} \\
 & Degree assortativity item & $\rho_{\mathcal{I}}$ &  & $Assort\text{-}I$\\
\bottomrule
\end{tabular}
\end{adjustbox}
\end{table}
\begin{table}[!b]
\caption{Dataset overall statistics and characteristic aggregated statistics (minimum and maximum values, mean, and standard deviation) on the sampled sub-datasets.}
\label{tab:dataset-chars}
\begin{adjustbox}{width=1\textwidth,center}
    \begin{tabular}{l|cccc|cccc|cccc}
    \toprule
        \multicolumn{1}{c}{} & \multicolumn{4}{c}{\textbf{Yelp2018}} & \multicolumn{4}{c}{\textbf{Amazon-Book}} & \multicolumn{4}{c}{\textbf{Gowalla}} \\ \cmidrule{2-13}
        \multicolumn{1}{c}{} & \multicolumn{4}{c}{\makecell[c]{\textit{\ul{Overall Statistics}}\\\textbf{Users:} 25,677 \quad \textbf{Items:} 25,815\\\textbf{Interactions:} 696,865}} & \multicolumn{4}{c}{\makecell[c]{\textit{\ul{Overall Statistics}}\\\textbf{Users:} 70,679 \quad \textbf{Items:} 24,915\\\textbf{Interactions:} 846,434}} & \multicolumn{4}{c}{\makecell[c]{\textit{\ul{Overall Statistics}}\\\textbf{Users:} 29,858 \quad \textbf{Items:} 40,981\\\textbf{Interactions:} 1,027,370}} \\ \cmidrule{2-13}
        \multicolumn{1}{c}{\textbf{Characteristics}} & \textbf{Min} & \textbf{Max} & \textbf{Mean} & \multicolumn{1}{c}{\textbf{Std}} & \textbf{Min} & \multicolumn{1}{c}{\textbf{Max}} & \textbf{Mean} & \multicolumn{1}{c}{\textbf{Std}} & \textbf{Min} & \multicolumn{1}{c}{\textbf{Max}} & \textbf{Mean} & \textbf{Std} \\
        \cmidrule{1-13}  
        $SpaceSize_{log}$ & 0.256 & 1.393 & 1.000 & 0.379 & 0.405 & 1.593 & 1.176 & 0.384 & 0.430 & 1.541 & 1.161 & 0.375 \\
        $Shape_{log}$ & 0.019 & 0.105 & 0.045 & 0.014 & 0.325 & 0.443 & 0.407 & 0.021 & -0.149 & -0.097 & -0.129 & 0.008 \\ 
        $Density_{log}$ & -3.699 & -2.693 & -3.219 & 0.358 & -3.902 & -2.896 & -3.497 & 0.365 & -3.889 & -2.876 & -3.380 & 0.363 \\ 
        $Gini\text{-}U$ & 0.443 & 0.508 & 0.486 & 0.008 & 0.384 & 0.499 & 0.459 & 0.023 & 0.462 & 0.512 & 0.491 & 0.007 \\ 
        $Gini\text{-}I$ & 0.500 & 0.609 & 0.575 & 0.019 & 0.518 & 0.618 & 0.586 & 0.018 & 0.437 & 0.502 & 0.478 & 0.008 \\ 
        $AvgDegree\text{-}U_{log}$ & 0.523 & 0.926 & 0.758 & 0.110 & 0.318 & 0.609 & 0.476 & 0.077 & 0.603 & 1.017 & 0.846 & 0.115 \\ 
        $AvgDegree\text{-}I_{log}$ & 0.565 & 0.955 & 0.804 & 0.098 & 0.682 & 1.043 & 0.883 & 0.096 & 0.487 & 0.888 & 0.717 & 0.109 \\ 
        $AvgClustC\text{-}U_{log}$ & -1.144 & -0.662 & -0.947 & 0.126 & -0.757 & -0.407 & -0.602 & 0.095 & -1.211 & -0.741 & -1.013 & 0.122 \\ 
        $AvgClustC\text{-}I_{log}$ & -1.092 & -0.652 & -0.922 & 0.105 & -1.124 & -0.751 & -0.967 & 0.099 & -1.080 & -0.614 & -0.881 & 0.124 \\ 
        $Assort\text{-}U$ & -0.051 & 0.235 & 0.021 & 0.035 & -0.041 & 0.533 & 0.052 & 0.074 & 0.042 & 0.544 & 0.188 & 0.071 \\ 
        $Assort\text{-}I$ & -0.002 & 0.237 & 0.067 & 0.037 & 0.000 & 0.842 & 0.443 & 0.264 & -0.037 & 0.161 & 0.021 & 0.028 \\ \bottomrule
    \end{tabular}
    \end{adjustbox}
\end{table}
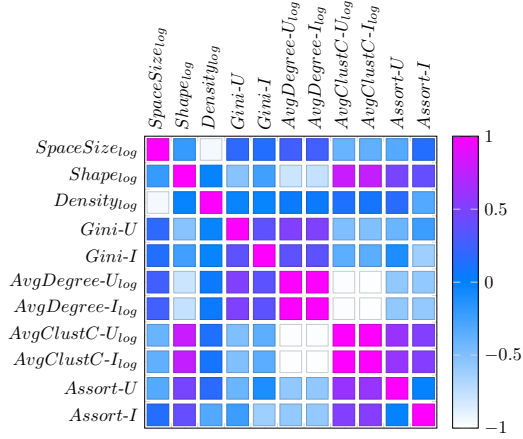
\begin{figure}[!ht]
\begin{adjustbox}{width=0.5\textwidth,center}
\begin{tikzpicture}
\begin{axis}[
    axis equal image, % We want a square grid, not a rectangular one
    scatter, % This activates the color mapping
    colormap/cool, % Choose the colormap
    colorbar, % Add a legend
    point meta min=-1,
    point meta max=1,
    grid=none, % Grid lines between the intervals
    minor tick num=1, % One minor tick per interval
    tickwidth=0pt, % Don't draw the major tick marks
    y dir=reverse, % Y increases downward
    xticklabel pos=right, % X axis labels go to the top
    xticklabel style={rotate=90},
    xtick={0,1,2,3,4,5,6,7,8,9,10},
    xticklabels={$SpaceSize_{log}$, $Shape_{log}$, $Density_{log}$, $Gini\text{-}U$, $Gini\text{-}I$, $AvgDegree\text{-}U_{log}$, $AvgDegree\text{-}I_{log}$, $AvgClustC\text{-}U_{log}$, $AvgClustC\text{-}I_{log}$, $Assort\text{-}U$, $Assort\text{-}I$},
    ytick={0,1,2,3,4,5,6,7,8,9,10},
    yticklabels={$SpaceSize_{log}$, $Shape_{log}$, $Density_{log}$, $Gini\text{-}U$, $Gini\text{-}I$, $AvgDegree\text{-}U_{log}$, $AvgDegree\text{-}I_{log}$, $AvgClustC\text{-}U_{log}$, $AvgClustC\text{-}I_{log}$, $Assort\text{-}U$, $Assort\text{-}I$},
    enlargelimits={abs=0.5}, % Add half a unit on all sides
    scatter/@pre marker code/.append code={% A bit of magic for scaling the circles in axis units
      \pgfplotstransformcoordinatex{sqrt(abs(\pgfplotspointmeta))}% Area scales with the square of the radius
      \scope[mark size=6, fill=mapped color]
    },
    scatter/@post marker code/.append code={%
      \endscope%
    }
]
\addplot +[
    point meta=explicit, % We'll provide values for the color and size
    mark=square*,
    only marks, % No lines between the points
    ] table [
    x expr={int(mod(\coordindex+0.01,11))}, % The position depends on the coordinate index, not the X or Y values
    y expr={int((\coordindex+0.01)/11))},
    meta=value
] {
X   Y   value

0	0	1.000000
0	1	-0.212315
0	2	-0.954520
0	3	0.180507
0	4	0.134404
0	5	0.255242
0	6	0.254337
0	7	-0.408143
0	8	-0.377079
0	9	-0.328141
0	10	0.143009
1	0	-0.212315
1	1	1.000000
1	2	-0.021688
1	3	-0.524974
1	4	-0.254467
1	5	-0.794564
1	6	-0.767826
1	7	0.783785
1	8	0.765346
1	9	0.469329
1	10	0.394994
2	0	-0.954520
2	1	-0.021688
2	2	1.000000
2	3	-0.031705
2	4	-0.027120
2	5	0.044572
2	6	0.045499
2	7	0.118263
2	8	0.084351
2	9	0.163524
2	10	-0.325265
3	0	0.180507
3	1	-0.524974
3	2	-0.031705
3	3	1.000000
3	4	0.371105
3	5	0.506309
3	6	0.497189
3	7	-0.497250
3	8	-0.503362
3	9	-0.415233
3	10	-0.228037
4	0	0.134404
4	1	-0.254467
4	2	-0.027120
4	3	0.371105
4	4	1.000000
4	5	0.361340
4	6	0.363311
4	7	-0.362226
4	8	-0.355434
4	9	-0.111091
4	10	-0.620275
5	0	0.255242
5	1	-0.794564
5	2	0.044572
5	3	0.506309
5	4	0.361340
5	5	1.000000
5	6	0.999082
5	7	-0.984254
5	8	-0.989383
5	9	-0.569820
5	10	-0.573792
6	0	0.254337
6	1	-0.767826
6	2	0.045499
6	3	0.497189
6	4	0.363311
6	5	0.999082
6	6	1.000000
6	7	-0.983229
6	8	-0.989941
6	9	-0.568128
6	10	-0.577563
7	0	-0.408143
7	1	0.783785
7	2	0.118263
7	3	-0.497250
7	4	-0.362226
7	5	-0.984254
7	6	-0.983229
7	7	1.000000
7	8	0.998075
7	9	0.598835
7	10	0.511454
8	0	-0.377079
8	1	0.765346
8	2	0.084351
8	3	-0.503362
8	4	-0.355434
8	5	-0.989383
8	6	-0.989941
8	7	0.998075
8	8	1.000000
8	9	0.594505
8	10	0.523383
9	0	-0.328141
9	1	0.469329
9	2	0.163524
9	3	-0.415233
9	4	-0.111091
9	5	-0.569820
9	6	-0.568128
9	7	0.598835
9	8	0.594505
9	9	1.000000
9	10	-0.013674
10	0	0.143009
10	1	0.394994
10	2	-0.325265
10	3	-0.228037
10	4	-0.620275
10	5	-0.573792
10	6	-0.577563
10	7	0.511454
10	8	0.523383
10	9	-0.013674
10	10	1.000000

};
\end{axis}
\end{tikzpicture}
\end{adjustbox}
\caption{Pearson correlation of the selected characteristics. Many values in $[-0.5, 0.5]$ indicate loosely correlated pairs. }
\label{fig:correlation}
\end{figure}

\subsubsection{Reproducibility} We perform the random subsampling strategy to split each sub-dataset into train and test (80\% and 20\%, respectively). Then, we retain the 10\% of the train as validation for the early stopping to avoid overfitting. To train LightGCN, DGCF, UltraGCN, and SVD-GCN, we fix their configurations (i.e., hyper-parameters and patience for the early stopping) to the best values according to the original papers, since our scope is not to fine-tune them. Finally, following the literature, we use the \textit{Recall@20} calculated on the validation for the early stopping, and evaluate the models by assessing the same metric on the test set. Codes, datasets, and configuration files to reproduce all the experiments are available at this link: ~\url{https://github.com/sisinflab/Graph-Characteristics}.

\subsection{Additional results for RQ1}
\label{app:add_exp_1}
Table \ref{tab:rq1_app} reports additional results for RQ1 on Amazon-Book, when considering the \textit{Recall@20} as recommendation metric (as done in Table \ref{tab:rq1}). Conversely, Table \ref{tab:rq1_ndcg} displays additional results for RQ1 when considering all three datasets for the \textit{nDCG@20} as recommendation metric.

\begin{table}[!h]
\caption{Additional results for RQ1 on Amazon-Book. The current table is to be interpreted the same way as Table \ref{tab:rq1}.}
\label{tab:rq1_app}
\begin{adjustbox}{width=0.6\textwidth,center}
    \centering
    \begin{tabular}{l|c|c|c|c}
    \toprule
        \multicolumn{1}{c}{\textbf{Characteristics}} & \multicolumn{1}{c}{\textbf{LightGCN}} & \multicolumn{1}{c}{\textbf{DGCF}} & \multicolumn{1}{c}{\textbf{UltraGCN}} & \multicolumn{1}{c}{\textbf{SVD-GCN}} \\ \cmidrule{1-5} 
        \rowcolor{gray} \ $R^{2}$(adj. $R^{2}$) & $0.953 (0.952)$ & $0.951 (0.950)$ & $0.800 (0.797)$& $0.964 (0.963)$
 \\ \
         $Constant$ & $0.088^{***}$&$0.079^{***}$&$0.056^{***}$&$0.112^{***}$
 \\ \
         $SpaceSize_{log}$ & $0.456^{***}$& $0.364^{***}$& $0.067^{}$& $0.433^{***}$
 \\ \
         $Shape_{log}$ &$-0.668^{***}$&$-0.598^{***}$&$-0.215^{}$&$-0.582^{***}$
 \\ \
         $Density_{log}$ & $0.546^{***}$& $0.453^{***}$& $0.141^{**}$&$0.541^{***}$
 \\ \
         $Gini\text{-}U$ &$-0.073^{}$&$-0.085^{}$&$-0.350^{**}$&$-0.042^{}$
 \\ \
         $Gini\text{-}I$ & $1.302^{***}$& $1.148^{***}$& $0.772^{***}$&$1.306^{***}$
 \\ \
         $AvgDegree\text{-}U_{log}$ &$1.336^{***}$&$1.116^{***}$&$0.316^{*}$&$1.265^{***}$
 \\ \
         $AvgDegree\text{-}I_{log}$ & $0.668^{***}$& $0.518^{***}$& $0.101^{}$& $0.683^{***}$
 \\ \
         $AvgClustC\text{-}U_{log}$ &$1.627^{***}$&$1.307^{***}$&$0.210^{}$&$1.617^{***}$
 \\ \
         $AvgClustC\text{-}I_{log}$ &$0.431^{***}$&$0.337^{***}$& $0.111^{}$& $0.354^{***}$
 \\ \
        $Assort\text{-}U$ &$0.031^{}$&$0.041^{*}$&$0.070^{***}$&$0.028^{}$
 \\ \
         $Assort\text{-}I$ &$-0.010^{}$& $-0.004^{}$& $0.025^{***}$& $-0.010^{*}$ \\ 
   \bottomrule
 \multicolumn{5}{l}{\textit{***p-value $\leq$ 0.001, **p-value $\leq$ 0.01, *p-value $\leq$ 0.05}}
    \end{tabular}
\end{adjustbox}
\end{table}
\begin{table}[!h]
\caption{Additional results for RQ1 on Yelp2018, Gowalla, and Amazon-Book when considering the \textit{nDCG@20} as accuracy metric. The current table is to be interpreted the same way as Table \ref{tab:rq1}, but \textit{nDCG@20} is the recommendation metric this time.}
\label{tab:rq1_ndcg}
    \centering
    \begin{adjustbox}{width=\textwidth,center}
    \begin{tabular}{l|ccc|ccc|ccc|ccc}
    \toprule
        \multicolumn{1}{c}{\multirow{2}{*}{\textbf{Characteristics}}} & \multicolumn{3}{c}{\textbf{LightGCN}} & \multicolumn{3}{c}{\textbf{DGCF}} & \multicolumn{3}{c}{\textbf{UltraGCN}} & \multicolumn{3}{c}{\textbf{SVD-GCN}} \\ \cmidrule{2-13}
        \multicolumn{1}{c}{} & \multicolumn{1}{c}{\textbf{Yelp2018}} & \textbf{Amazon-Book} & \multicolumn{1}{c}{\textbf{Gowalla}} & \textbf{Yelp2018} & \textbf{Amazon-Book} & \multicolumn{1}{c}{\textbf{Gowalla}} & \textbf{Yelp2018} & \textbf{Amazon-Book} & \multicolumn{1}{c}{\textbf{Gowalla}} & \textbf{Yelp2018} & \textbf{Amazon-Book} & \textbf{Gowalla} \\ \cmidrule{1-13}
        \rowcolor{gray} $R^{2}$(adj. $R^{2}$) & $0.971 (0.970)$& $0.947 (0.946)$& $0.978 (0.977)$& $0.972 (0.971)$& $0.945 (0.944)$& $0.980 (0.979)$& $0.966 (0.966)$& $0.748 (0.743)$& $0.865 (0.863)$& $0.982 (0.982)$& $0.958 (0.957)$& $0.980 (0.980)$ \\ \
         $Constant$ & $0.048^{***}$&$0.043^{***}$&$0.068^{***}$&$0.043^{***}$&$0.039^{***}$&$0.061^{***}$&$0.028^{***}$&$0.025^{***}$&$0.034^{***}$&$0.056^{***}$&$0.054^{***}$&$0.076^{***}$ \\ \
         $SpaceSize_{log}$ & $0.054^{***}$& $0.286^{***}$& $0.142^{***}$& $0.086^{***}$& $0.229^{***}$& $0.170^{***}$& $-0.017^{**}$& $0.054^{}$& $0.216^{***}$& $0.062^{***}$& $0.290^{***}$& $0.107^{***}$ \\ \
         $Shape_{log}$ & $-0.191^{***}$&$-0.447^{***}$&$-0.168^{*}$&$-0.185^{***}$&$-0.389^{***}$&$-0.134^{}$&$0.062^{}$&$-0.154^{}$&$-0.019^{}$&$-0.159^{**}$&$-0.417^{***}$&$-0.193^{*}$ \\ \
         $Density_{log}$ &  $0.115^{***}$& $0.332^{***}$& $0.199^{***}$& $0.140^{***}$& $0.274^{***}$& $0.218^{***}$& $0.024^{***}$& $0.090^{**}$& $0.213^{***}$& $0.131^{***}$& $0.345^{***}$& $0.172^{***}$
 \\ \
         $Gini\text{-}U$ & $0.140^{**}$&$-0.106^{}$&$-0.034^{}$&$0.046^{}$&$-0.105^{}$&$-0.085^{}$&$0.060^{}$&$-0.194^{**}$&$-0.547^{***}$&$0.050^{}$&$-0.140^{*}$&$0.012^{}$
 \\ \
         $Gini\text{-}I$ & $0.678^{***}$& $0.720^{***}$& $0.438^{***}$& $0.588^{***}$& $0.624^{***}$& $0.368^{***}$& $0.264^{***}$& $0.393^{***}$& $-0.145^{}$& $0.561^{***}$& $0.722^{***}$& $0.482^{***}$
 \\ 
         $AvgDegree\text{-}U_{log}$ & $0.264^{***}$&$0.841^{***}$&$0.425^{***}$&$0.319^{***}$&$0.698^{***}$&$0.455^{***}$&$-0.024^{}$&$0.221^{*}$&$0.439^{***}$&$0.273^{***}$&$0.843^{***}$&$0.375^{***}$
 \\ \
         $AvgDegree\text{-}I_{log}$ & $0.073^{*}$& $0.394^{***}$& $0.257^{***}$& $0.134^{***}$& $0.309^{***}$& $0.322^{***}$& $0.038^{}$& $0.068^{}$& $0.420^{***}$& $0.113^{***}$& $0.426^{***}$& $0.182^{***}$
 \\ \
         $AvgClustC\text{-}U_{log}$ & $0.405^{***}$&$1.050^{***}$&$0.514^{***}$&$0.466^{***}$&$0.843^{***}$&$0.551^{***}$&$-0.006^{}$&$0.167^{}$&$0.461^{***}$&$0.441^{***}$&$1.094^{***}$&$0.443^{***}$
 \\ \
         $AvgClustC\text{-}I_{log}$ & $0.024^{}$& $0.205^{***}$& $0.193^{}$& $0.070^{}$& $0.163^{**}$& $0.258^{**}$& $0.038^{}$& $0.062^{}$& $0.461^{***}$& $0.016^{}$& $0.174^{**}$& $0.146^{}$
 \\ \
          $Assort\text{-}U$ & $0.058^{***}$&$0.018^{}$&$0.020^{**}$&$0.055^{***}$&$0.022^{*}$&$0.014^{*}$&$0.093^{***}$&$0.045^{***}$&$0.005^{}$&$0.048^{***}$&$0.016^{}$&$0.008^{}$
 \\ \
         $Assort\text{-}I$ & $-0.029^{*}$& $-0.005^{}$& $-0.034^{}$& $-0.029^{*}$& $-0.002^{}$& $-0.046^{}$& $0.005^{}$& $0.016^{***}$& $-0.113^{***}$& $-0.028^{*}$& $-0.006^{*}$& $-0.074^{**}$
 \\ 
   \bottomrule
 \multicolumn{13}{l}{\textit{***p-value $\leq$ 0.001, **p-value $\leq$ 0.01, *p-value $\leq$ 0.05}}
    \end{tabular}
    \end{adjustbox}
\end{table}

\subsection{Intuition for RQ2}
\label{app:intuition}

Figure \ref{fig:scale-free} displays the relation (i.e., the black points) between the probability distribution of node degrees in the original graph and their degree values on the Gowalla dataset. As evident, high-degree nodes are less popular than low-degree ones, and this resembles the tendency of real-world networks to be \textit{scale-free}~\cite{PhysRevE.67.026112}. To be more precise, the actual degree probability distribution approximates neither the \textit{power-law} (i.e., representing \textit{scale-free} networks, in green), nor the \textit{exponential} function (i.e., in red), but it would be well-approximated by a function in-between. This suggests that the high-degree nodes are even less frequent than they usually are in \textit{scale-free} networks.  

The figure helps to re-interpret the impact of node- and edge-dropout. While node-dropout works by removing nodes (and all the edges connected to them), edge-dropout eliminates edges and the consequently-disconnected nodes. Let us consider their worst-case scenarios. For node-dropout, it would be to drop many high-degree nodes from the graph. Whereas, when considering edge-dropout, it would be to drop all the edges connected to several nodes and thus disconnect them from the graph. 

This intuition drove us towards stating that, on averagely, node-dropout has the potential to drop larger portions of the user-item graph than edge-dropout. Indeed, this might undermine the goodness of the explanations produced by our explanatory framework. The assumption further motivates the strategy we adopted to generate the sub-datasets in RQ1, where we performed both node- and edge-dropout by uniformly selecting one of them for each sampled sub-dataset in order not to bias the procedure towards either node- or edge-dropout (see again Section \ref{dataset-generation}).

\input{figures/scale_free.tex}

\subsection{Additional results for RQ2}
\label{app:add_exp_2}
Table \ref{tab:rq2_app} reports additional results for RQ2. 
\begin{table}[!h]
\caption{Additional results for RQ2 on DGCF and UltraGCN. The current table is to be interpreted the same way as Table \ref{tab:rq2}.}
\label{tab:rq2_app}
\begin{adjustbox}{width=\textwidth,center}
    \centering
    \begin{tabular}{l|cc|cc|cc|cc}
    \toprule
         \multicolumn{1}{c}{} & \multicolumn{2}{c}{\textbf{Node drop} \tikzcircle[fill=black]{2.5pt} \tikzcircle[fill=black]{2.5pt} \tikzcircle[fill=black]{2.5pt} \quad
         \textbf{Edge drop} \tikzcircle[fill=white]{2.5pt} \tikzcircle[fill=white]{2.5pt} \tikzcircle[fill=white]{2.5pt}} & \multicolumn{2}{c}{\textbf{Node drop} \tikzcircle[fill=black]{2.5pt} \tikzcircle[fill=black]{2.5pt} \tikzcircle[fill=white]{2.5pt} \quad
         \textbf{Edge drop} \tikzcircle[fill=black]{2.5pt} \tikzcircle[fill=white]{2.5pt} \tikzcircle[fill=white]{2.5pt}} & \multicolumn{2}{c}{\textbf{Node drop} \tikzcircle[fill=black]{2.5pt} \tikzcircle[fill=white]{2.5pt} \tikzcircle[fill=white]{2.5pt} \quad
         \textbf{Edge drop} \tikzcircle[fill=black]{2.5pt} \tikzcircle[fill=black]{2.5pt} \tikzcircle[fill=white]{2.5pt}} & \multicolumn{2}{c}{\textbf{Node drop} \tikzcircle[fill=white]{2.5pt} \tikzcircle[fill=white]{2.5pt} \tikzcircle[fill=white]{2.5pt} \quad
         \textbf{Edge drop} \tikzcircle[fill=black]{2.5pt} \tikzcircle[fill=black]{2.5pt} \tikzcircle[fill=black]{2.5pt}} \\ \cmidrule{2-9}
         \multicolumn{1}{c}{} & \multicolumn{2}{c}{\makecell[c]{\textit{\ul{Average Sampling Statistics}}\\\textbf{Users:} 5,828 \quad \textbf{Items:} 7,887 \\ \textbf{Interactions:} 45,620}} & \multicolumn{2}{c}{\makecell[c]{\textit{\ul{Average Sampling Statistics}}\\\textbf{Users:} 12,744 \quad \textbf{Items:} 17,229 \\ \textbf{Interactions:} 97,785}} & \multicolumn{2}{c}{\makecell[c]{\textit{\ul{Average Sampling Statistics}}\\\textbf{Users:} 21,730 \quad \textbf{Items:} 29,316 \\ \textbf{Interactions:} 160,919\\}} & \multicolumn{2}{c}{\makecell[c]{\textit{\ul{Average Sampling Statistics}}\\\textbf{Users:} 28,526 \quad \textbf{Items:} 38,467 \\ \textbf{Interactions:} 209,659}} \\ \cmidrule{2-9} 
         \multicolumn{1}{c}{\textbf{Characteristics}} & \multicolumn{1}{c}{\textbf{DGCF}} & \multicolumn{1}{c}{\textbf{UltraGCN}} & \multicolumn{1}{c}{\textbf{DGCF}} & \multicolumn{1}{c}{\textbf{UltraGCN}} & \multicolumn{1}{c}{\textbf{DGCF}} & \multicolumn{1}{c}{\textbf{DGCF}} & \multicolumn{1}{c}{\textbf{UltraGCN}} \\ 
         \cmidrule{1-9} 
         \rowcolor{gray} $R^{2}$(adj. $R^{2}$) &$0.888 (0.884)$&$0.597 (0.583)$ &$0.973 (0.972)$&$0.833 (0.827)$&$0.988 (0.988)$&$0.883 (0.879)$&$0.994 (0.994)$&$0.599 (0.584)$ \\
        $Constant$ &$0.162^{***}$&$0.091^{***}$&$0.131^{***}$&$0.074^{***}$&$0.085^{***}$&$0.051^{***}$&$0.051^{***}$&$0.034^{***}$ \\
        $SpaceSize_{log}$ &$0.130^{**}$&$0.175^{*}$&$0.192^{***}$&$0.358^{***}$&$0.264^{***}$&$0.337^{***}$&$-0.188^{***}$&$1.644^{***}$ \\
        $Shape_{log}$ &$-0.109^{}$&$-0.096^{}$&$-0.232^{*}$&$-0.022^{}$&$-0.136^{}$&$-0.029^{}$&$-0.011^{}$&$1.055^{**}$ \\
        $Density_{log}$ &$-0.005^{}$&$0.394^{***}$&$0.294^{***}$&$0.355^{***}$&$0.351^{***}$&$0.340^{***}$&$0.168^{***}$&$0.185^{}$ \\
        $Gini\text{-}U$ &$-0.136^{}$&$-0.824^{***}$&$0.083^{}$&$-0.880^{***}$&$0.130^{}$&$-0.775^{**}$&$0.200^{*}$&$-0.961^{}$ \\
        $Gini\text{-}I$ &$0.756^{***}$&$-0.152^{}$&$0.717^{***}$&$-0.248^{}$&$0.668^{***}$&$-0.333^{}$&$0.264^{**}$&$-0.732^{}$ \\
        $AvgDegree\text{-}U_{log}$ &$0.179^{**}$&$0.617^{***}$&$0.601^{***}$&$0.723^{***}$&$0.684^{***}$&$0.692^{***}$&$-0.014^{}$&$1.302^{***}$\\
        $AvgDegree\text{-}I_{log}$ &$0.070^{}$&$0.521^{**}$&$0.369^{***}$&$0.702^{***}$&$0.547^{***}$&$0.663^{***}$&$-0.025^{}$&$2.356^{***}$ \\
        $AvgClustC\text{-}U_{log}$ &$0.362^{***}$&$0.619^{**}$&$0.715^{***}$&$0.641^{**}$&$0.706^{***}$&$0.481^{}$&$0.001^{}$&$1.099^{**}$\\
        $AvgClustC\text{-}I_{log}$ &$-0.080^{}$&$0.412^{}$&$0.252^{}$&$0.903^{**}$&$0.571^{**}$&$0.962^{**}$&$-0.083^{}$&$2.415^{***}$ \\
        $Assort\text{-}U$ &$-0.001^{}$&$0.001^{}$&$0.011^{}$&$0.000^{}$&$0.001^{}$&$0.012^{}$&$0.008^{}$&$0.064^{}$ \\
        $Assort\text{-}I$ &$-0.050^{}$&$-0.111^{**}$&$0.002^{}$&$-0.151^{**}$&$0.002^{}$&$-0.086^{}$&$-0.032^{}$&$-0.141^{}$
         \\
    \bottomrule
    \multicolumn{9}{l}{\textit{***p-value $\leq$ 0.001, **p-value $\leq$ 0.01, *p-value $\leq$ 0.05}}
    \end{tabular}
    \end{adjustbox}
\end{table}

% For bibLaTeX users:
% \printbibliography

\end{document}